\documentclass[journal]{IEEEtran}
\usepackage{xcolor,soul,framed} 
\colorlet{shadecolor}{yellow}
\usepackage[pdftex]{graphicx}
\graphicspath{{../pdf/}{../jpeg/}}
\DeclareGraphicsExtensions{.pdf,.jpeg,.png}
\usepackage[cmex10]{amsmath}
\usepackage{array}
\usepackage{amssymb}
\usepackage{mdwmath}

\usepackage{mdwtab}
\usepackage{eqparbox}
\usepackage{url}
\usepackage{enumerate}
\usepackage{enumitem}
\usepackage{amsfonts}
\usepackage[english]{babel}
\usepackage[utf8]{inputenc}
\usepackage[noend]{algpseudocode}
\usepackage{mathtools}
\usepackage{nomencl}
\usepackage{blindtext}
\usepackage{amsmath}
\usepackage[colorinlistoftodos]{todonotes}
\usepackage{color}
\usepackage{color,soul}
\usepackage{xcolor}
\usepackage{multirow}
\usepackage{graphicx}
\usepackage{epstopdf}
\usepackage[T1]{fontenc}
\usepackage{textcase}
\usepackage{array}
\usepackage{mdwmath}
\usepackage{mdwtab}
\usepackage{eqparbox}
\usepackage{enumerate}
\usepackage{bm}
\usepackage{amsfonts}
\usepackage[ruled,lined,linesnumbered]{algorithm2e}
\usepackage{booktabs}
\usepackage{comment}
\usepackage{amsmath}
\usepackage{mathtools}
\usepackage{caption}
\usepackage{subcaption}
\usepackage{tabularx}
\usepackage{hyperref}
\hypersetup{
    colorlinks=true,
    linkcolor=blue,
    filecolor=magenta,      
    urlcolor=cyan,
}
\usepackage{cleveref}

\usepackage{subcaption}
\usepackage{caption}
\usepackage{mathrsfs}
\usepackage{booktabs}
\usepackage{siunitx}
\usepackage{pythonhighlight}
\usepackage{textcomp}
\usepackage{listings}
\usepackage{float}   

\usepackage{cite}
\usepackage{url}
\usepackage{setspace}
\usepackage{amsthm}
\usepackage{nomencl}
\usepackage{footnote}
\usepackage{threeparttable}
\usepackage{tikz}
\usepackage{tabularx}
\usepackage{enumitem}

\newcommand{\dg}{\operatorname{dg}}
\usepackage{cancel}

\usepackage{pifont}
\newcommand{\cmark}{\ding{51}}%
\newcommand{\xmark}{\ding{55}}%

\usepackage{newfloat}
\usepackage{listings}

\usepackage{booktabs}
\usepackage{siunitx}
\usepackage{adjustbox}

\usepackage[most]{tcolorbox}

\DeclareCaptionStyle{ruled}{labelfont=normalfont,labelsep=colon,strut=off} 
\floatstyle{ruled}
\newfloat{listing}{tb}{lst}{}
\floatname{listing}{Listing}
\usepackage{graphicx}
\usepackage{placeins}

\usepackage{wrapfig}

\makenomenclature
\usepackage{etoolbox}
\renewcommand\nomgroup[1]{%
  \item[\bfseries
  \ifstrequal{#1}{A}{Abbreviations}{%
  \ifstrequal{#1}{V}{Vectors, Matrices, and Sets}{%
  \ifstrequal{#1}{O}{Operators}{%
  \ifstrequal{#1}{S}{Operators, Sets, and Symbols}{}}}}%
]}

\newtheorem{theorem}{Theorem}
\newtheorem{proposition}{Proposition}

\newtheorem{definition}{Definition}

\tikzset{square arrow/.style={to path={-- ++(0,-.25) -| (\tikztotarget)}}}

\usepackage{xcolor,cite,etoolbox}
\usepackage{tabularx}
\makeatletter 
\pretocmd\@bibitem{\color{black}\csname keycolor#1\endcsname}{}{\fail}

\usepackage{etoolbox}
\makeatletter
\patchcmd{\@makecaption}
  {\scshape}
  {}
  {}
  {}
\makeatother

\title{Structured Differentiable Optimization for Efficient Decision-focused Learning in Power Systems} 

\author{Wangkun Xu,~\IEEEmembership{Member,~IEEE},
    and Fei Teng,~\IEEEmembership{Senior Member,~IEEE}\\
}
\begin{document}
\setlength{\textfloatsep}{0.1pt}
\renewcommand{\baselinestretch}{1}
\markboth{Submitted to IEEE}%
{Shell \MakeLowercase{\emph{et al.}}: Bare Demo of IEEEtran.cls for IEEE Journals}
\maketitle

\begin{abstract}
    Decision-focused learning (DfL) trains forecasting models to align downstream decision consequences, such as power-system operating costs. However, its application to realistic power networks is limited by the need to repeatedly solve and differentiate large optimization problems during training. This paper presents DiffAPQP, a solver-flexible framework and open-source Python package for scalable DfL with affine-parametric quadratic programs. To accelerate the forward pass, DiffAPQP automatically canonicalizes quadratic power-system models written in CVXPY into a differentiation-ready representation and takes advantage of the repetitive solving structure through solver warm-start and solver-data update during training. For the backward pass acceleration, we establish the equivalence between differentiation through the full KKT system and a reduced system obtained by eliminating inactive inequality constraints. For training losses depending solely on the optimal value, we further derive an envelope-theorem-based gradient that avoids solving an adjoint KKT system, resulting in eligible backward time.
    To our knowledge, this work presents the first solver-based end-to-end DfL demonstration on the IEEE 118-bus system with a 24-hour coupled economic-dispatch and redispatch horizon. 
    Under matched SCS and Clarabel backends on a Linux machine, DiffAPQP achieves $2.27\times$--$3.58\times$ closed-loop and $3.62\times$--$4.38\times$ counterfactual end-to-end DfL training speedups over CvxpyLayers. The best solver configurations increase these speedups to $3.91\times$ (from 38.65 to 9.55 min/epoch) and $6.39\times$ (from 10.73 to 1.68 min/epoch), respectively. Additionally, DiffAPQP reduces peak memory usage by approximately $50\%$, while keeping similar operating costs as CvxpyLayers\footnote{DiffAPQP is available at \url{https://github.com/xuwkk/diffapqp} and can be installed by \texttt{pip install diffapqp}. The paper has an online companion with theoretical proofs, package design\&usage, and extra results at \url{}; all experiments of this paper can be found at \url{https://github.com/xuwkk/diffapqp_power_systems}.}.
\end{abstract}


\section{Introduction}

\subsection{Background}

Forecasts of renewable generation, load demand, and electricity prices are fundamental inputs to power system operational decision
 making. However, recent studies have shown that higher statistical forecasting accuracy does not necessarily translate into better operational decisions. For instance, in renewable generation forecasting, models trained using standard statistical metrics such as mean squared error (MSE) may fail to capture the asymmetric operational consequences of over- and under-forecasting. Moreover, due to the presence of physical and operational constraints in power system optimization, the relationship between forecast errors and operating costs is typically nonlinear and context-dependent, and therefore cannot be adequately represented by simple error-to-cost mapping \cite{zhang2022cost}. Similar observations have also emerged in practice. In the California ISO system, system operators may deliberately adjust load forecasts to improve downstream operational feasibility, such as increasing available ramping capacity  \cite{morales2023prescribing}. 

\subsection{Decision-focused Learning in Power Systems}

To better align forecasting with downstream operational outcomes, decision-focused learning (DfL) has recently been introduced into power system applications. In DfL, the downstream optimization problem is explicitly incorporated into the training process of the forecaster \cite{zhang2026decision}. Depending on the structure of the forecaster and the downstream optimization problem, DfL can be implemented through mathematical-programming-based or gradient-based approaches \cite{xu2025lapso}. The former typically casts DfL as a bilevel optimization problem, whose exact solution often requires restrictive assumptions, such as convex operational optimization and simple forecasting models, leading to computationally demanding mixed-integer reformulations \cite{chen2022feature, chen2024towards}. This paper instead considers neural network (NN)-based forecasters trained on large-scale datasets using stochastic gradient descent (SGD), for which scalable gradient-based DfL training algorithms are needed.

\begin{table}[t]
    \footnotesize
    \caption{Review on DfL learning in power systems with neural forecaster + optimization(s) structures.}
    \begin{tabularx}{\linewidth}{p{0.55cm}p{1.8cm}p{1.0cm}X}
    \toprule
    Ref. & Test system & Forecast & Operation (optimization class) \\
    \midrule
    \cite{donti2017task} & Grid aggregator (No network) & Load & Redispatch (Stochastic LP) \\\hline
    \cite{han2021task} & IEEE 14-Bus & Load &  Redispatch (Stochastic LP) \\\hline
    \cite{vohra2023end} & IEEE 6-Bus & Wind \& Solar & Dispatch / Redispatch (LPs) \\\hline
     \cite{zhang2024toward} & VPP (No network) & Wind & Dispatch / Redispatch (LPs) \\\hline
     \cite{wahdany2023more} & IEEE 6-Bus & Wind & Dispatch (LP) \\\hline
     \cite{xu2024e2e} & IEEE 14-Bus & Load & Dispatch / redispatch (LP) \\\hline
     \cite{xu2024task} & IEEE 14-Bus & Load & Dispatch / redispatch (QP) \\\hline
     \cite{stratigakos2025decision} & Grid aggregator (No network) & Wind & Redispatch (Stochastic LP) \\\hline
     \cite{paredes2025participation} & BESS (no network) & Price, etc & Reserve market clearing / real-time correction (LPs) \\
     \bottomrule
\end{tabularx}
    \label{tab:dfl_power_system_review}
\end{table}

As summarized in Table~\ref{tab:dfl_power_system_review}, a standard DfL pipeline in power systems consists of three key components: a neural forecaster for renewable generation, load, and/or market price prediction; one or more optimization problems representing operational decision-making tasks, such as dispatch and redispatch; and a task-oriented loss function that evaluates the quality of the resulting decisions. Compared with conventional NN training, DfL introduces an optimization problem into the computational graph. Therefore, the forward pass requires solving the downstream optimization problem(s), typically through an iterative numerical algorithm, while the backward pass requires differentiating through the solution mapping of this optimization problem. The differentiable optimization (DiffOpt) can be achieved either by backpropagating through the unrolled solver iterations, known as \emph{unrolled differentiation} \cite{kotary2023backpropagation}, or by differentiating the optimality conditions around the solution, referred to as \emph{implicit differentiation} \cite{blondel2024elements}.

Most existing DfL studies in power systems implement this pipeline using open-source DiffOpt packages, particularly CvxpyLayers \cite{agrawal2019differentiating}. CvxpyLayers builds on the high-level optimization modeling framework CVXPY \cite{diamond2016cvxpy}, allowing practitioners to formulate power system optimization problems in a descriptive and engineering-friendly syntax. The resulting problem is then canonicalized and solved in the forward pass as a \emph{standard} conic program using solvers such as SCS \cite{o2021operator} and Clarabel \cite{gaulart2024clarabel}. Gradients for the backward pass are obtained through implicit differentiation of the associated Karush--Kuhn--Tucker (KKT) optimality conditions \cite{amos2019differentiable}.

Because both forward optimization and backward differentiation are computationally demanding, existing CvxpyLayers-based DfL studies in power systems have predominantly considered aggregated or small-scale systems, shortened scheduling horizons, or simplified network constraints (Table~\ref{tab:dfl_power_system_review}). Therefore, evidence of end-to-end DfL performance at realistic operational scales remains limited. Moreover, when the operational model is formulated as a linear program (LP), the singular KKT system can result in non-unique gradient, hindering stable convergence of DfL training \cite{paredes2025participation, wilder2019melding}. 

\subsection{Review on Existing Efficient DfL Training Strategies}

To improve training efficiency, one research line focuses on first learning a surrogate NN model for the optimization layer, including direct link forecast error to operational error \cite{zhang2022cost,li2018forecasting} or link the forecasted parameters with the optimal decisions \cite{kotary2024learning}. Although this can significantly improve the DfL training efficiency as the trained model can be treated as a regular NN layer and no iterative solution or implicit differentiation is required, training such surrogate requires collecting extra dataset and the generality on out-of-distribution operational scenarios cannot be easily guaranteed. Therefore, this paper focuses on DfL training where the exact optimization is embedded as a layer.

Another line of work focuses on reducing the KKT system used for differentiating the solution map of quadratic programs. For example, BPQP \cite{pan2024bpqp} and dQP \cite{magoon2026differentiation} exploit the active set at the optimum so that the backward pass only involves a reduced equality-constrained KKT system, and dQP further decouples this differentiation step from the forward solver, allowing the use of black-box QP solvers. However, these methods operate on QPs already expressed in standard matrix form. Therefore, the obtained gradients are with respect to canonical QP data, such as the objective and constraint matrices/vectors, rather than directly with respect to application-level parameters such as renewable generation forecasts. Recovering such gradients still requires an additional differentiable mapping from the original modeling parameters to the canonical QP data.

\subsection{Contributions}

\begin{table}[t]
    \centering
    \footnotesize
    \caption{Functionality overview of the proposed DiffAPQP, compared to CvxpyLayers.}
    \begin{threeparttable}
    \begin{tabularx}{\linewidth}{r|X|X}
        & \textbf{DiffAPQP} & \textbf{CvxpyLayers} \\\hline\hline
        \textbf{Class of Program} & APQP & Disciplined CP  \\
        \textbf{Canonicalization} & \cmark & \cmark \\
        \textbf{Solver Backend} & All QP-compatible solvers in CVXPY & Limited to Clarabel, SCS, and ECOS  \\
        \textbf{Warm Start}$^\star$ & \cmark & \xmark \\
        \textbf{Solver Update}$^\star$ & \cmark & \xmark \\
        \textbf{Parallel Forw. Solution} & \cmark & \cmark  \\
        \textbf{Reduced Backw. Diff.} & \cmark & \xmark \\
        \textbf{``Free'' Backw. Diff.} & \cmark$^\diamond$ & \xmark
    \end{tabularx}
    \begin{tablenotes}
      \item[$\star$] Currently supports OSQP, SCS, and QPALM, which have native warm-start and update APIs. The remaining solvers apply the CVXPY default warm-start settings.
      \item[$\diamond$] Only for value function-based training loss. 
      \end{tablenotes}
    \end{threeparttable}
    \label{tab:compare}
\end{table}

We identify two underexploited structures in existing DfL pipelines: forecast parameters enter the operational model through fixed affine mappings, and training process repeatedly solves the same optimization template with different parameter values. To explore the structure, We restrict the scope into quadratic program (QP)-based power system optimizations, and propose differentiable affine parametric quadratic program (DiffAPQP) layer. Compared to CvxpyLayers, this framework has the following properties or advantages (Table~\ref{tab:compare}): 

\underline{Methodologically}, \textbf{I).} in the forward pass, a dedicated automatic canonicalization is implemented so that no manual transformation to standard QP is needed and explicit gradients with respect to power system forecast parameters are available after backward propagation. \textbf{II).} As the power system optimization template is fixed during training across the entire dataset and epochs, we design solver-agnostic warm-start and solver-data update strategy on work space/factorization cache of previous solve, which significantly improve the forward pass efficiency.

\underline{Theoretically}, \textbf{III).} in the backward pass, we prove the equivalence of the implicit differentiation between full and a reduced equality-constrained QP KKT equilibrium for both QP and LP settings. This allows to solve a much smaller linear symmetric system. \textbf{IV).} We then derive a new backpropagation strategy \emph{without} differentiation for a value-function-induced training loss. 

\underline{Practically}, \textbf{V).} the proposed method disentangles forward and backward passes, meaning that it supports compatible continuous solver backends independently of the selected backward method. \textbf{VI).} At last, a Python package is open-sourced, which can automatically convert \emph{any} APQP power system optimization compiled by CVXPY, into differentiable optimization layer, that can be integrated with PyTorch with all the above features supported by APIs. 


\section{Preliminaries on the Decision-focused Learning in Power Systems}

\subsection{Sequential Chain of Forecast and Optimization}\label{sec:sequential_chain}

This paper considers a power-system decision chain consisting of a forecasting task followed by a power-system operation (PSO) problem \cite{conejo2018power}. The forecaster is denoted as $f(x,\theta)$ which takes contextual information $x$ such as weather, calendar information, and previous profile as input. In a supervised setting, the forecaster is trained to minimize the mean squared error (MSE) loss to the true label $y$ and the training can be denoted as $\min_\theta \ell_{AbL}(\theta) := \frac{1}{2N_\mathcal{D}}\sum_{(x,y)\in\mathcal{D}} \|y-f(x,\theta)\|^2$ where $\mathcal{D}$ is the training dataset with $N_\mathcal{D}$ samples. In addition, $f(\cdot,\theta)$ is considered as a neural network (NN) and stochastic gradient descent method (SGD) or its variant is used to update the parameter $\theta$ batch-wisely \cite{lecun2015deep}. Because the training objective is to minimize the forecaster error (or equivalently maximize the forecast accuracy), it is referred to as \emph{accuracy-based learning} (AbL).

When multiple profiles are forecast, such as load, renewable generation, reserve requirements, and market prices, the forecasts are collected in $\hat y_{1:N_p}$. Then the forecasts are fed into a power system operational model, which can be denoted as the following parametric program \cite{xu2024e2e},
\begin{equation}\label{eq:ori_prob}
    \begin{aligned}
       \min_{z_{1:N_d}} \;& f_0(z_{1:N_d},\hat{y}_{1:N_p}) \\
        \text{s.t.} \;& h_k(z_{1:N_d},\hat{y}_{1:N_p})= 0, \; \forall k=1,\cdots,N_{eq} \\
        & f_j(z_{1:N_d},\hat{y}_{1:N_p})\leq 0, \; \forall j=1,\cdots,N_{in}
    \end{aligned}
\end{equation}
where $N_d, N_p, N_{eq}$ and $N_{in}$ are the numbers of decision variables, parameters, equality constraints, and inequality constraints, respectively. For example, it can represent economic dispatch (ED) where $\hat{y}$ represents the forecasted load and renewable while $z$ represents dispatch and reserve plans. Moreover, in the energy market setting, it represents a bidding problem where $\hat{y}$ can include the day-ahead or balancing market prices. As each element is individually denoted by indices, \eqref{eq:ori_prob} represents the optimization followed by its natural \emph{physical} meanings, which is referred to as \emph{domain specific language}. A crucial step before invoking the optimization solver backend on \eqref{eq:ori_prob} is \emph{canonicalization}, in which \eqref{eq:ori_prob} is reformulated into the standard form required by the chosen solver. Some modeling software, such as CVXPY, can do this compilation automatically \cite{diamond2016cvxpy}. 

In the following discussion, the decision variables and parameters will be compactly denoted as $z$ and $\hat{y}$, respectively, and the optimum is denoted as $\hat{z}^\star$. Moreover, \eqref{eq:ori_prob} can be denoted as an \emph{implicit} function $\hat{z}^\star = \mathcal{O}(\hat{y})$ and the value function of \eqref{eq:ori_prob} is represented as $\alpha(\hat{y}) := f_0(\hat{z}^\star,\hat{y})$.

\subsection{Decision-focused Learning}

\begin{figure}[t]
     \centering
     \begin{subfigure}[b]{0.35\textwidth}
         \centering
         \includegraphics[width=\textwidth, trim={0.5cm 0 2.0cm 0},clip]{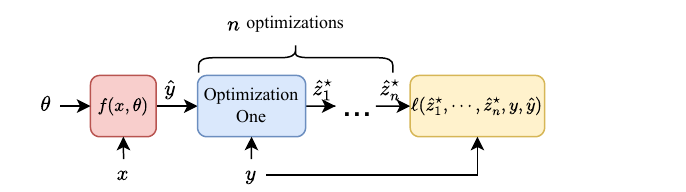}
         \caption{General DfL framework.}
         \label{fig:dfl_general}
     \end{subfigure}
     \hfill
     \begin{subfigure}[b]{0.35\textwidth}
         \centering
         \includegraphics[width=\textwidth,trim={0.5cm 0 2.0cm 0},clip]{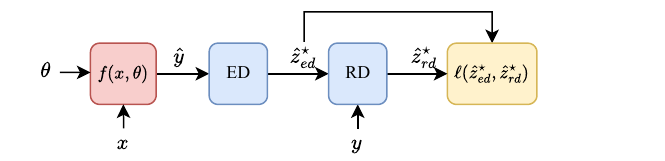}
         \caption{Closed-loop DfL for forecast-ED-RD decision makings.}
         \label{fig:dfl_closedloop}
     \end{subfigure}
     \hfill
     \begin{subfigure}[b]{0.35\textwidth}
         \centering
         \includegraphics[width=\textwidth,trim={0.5cm 0 0.0cm 0},clip]{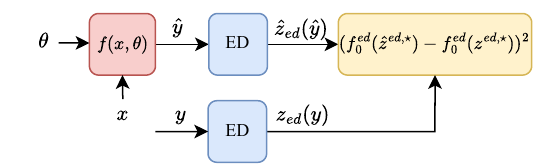}
         \caption{Counterfactual DfL for forecast-ED decision makings.}
         \label{fig:dfl_counterfactual}
     \end{subfigure}
        \caption{Illustration on DfL framework. Only the forward passes are shown.}
        \label{fig:dfl}
\end{figure}

To align the forecaster with the outcome of downstream power system optimizations, \emph{decision-focused learning} (DfL) is proposed to train the forecaster with optimizations in a closed-loop \cite{zhang2026decision}. From a mathematical program point of view, DfL is formulated as bi-level optimization problem \cite{xu2025lapso},
\begin{equation}\label{eq:dfl_general}
    \begin{aligned}
    \min_\theta ~~
    & \textstyle \frac{1}{N_\mathcal{D}} \sum_{(x,y)\in\mathcal{D}}
    \ell(\hat{z}^{1,\star},\ldots,\hat{z}^{n,\star},y,\hat{y}) \\
    \text{s.t.} ~~
    & \left\{
    \begin{aligned}
    & \hat{z}^{n,\star} = \mathcal{O}^n(\hat{z}^{n-1,\star}) \\
    & \hspace{1.5cm} \vdots \\
    & \hat{z}^{1,\star} = \mathcal{O}^1(\hat{y}), \; \hat{y} = f(x,\theta)
    \end{aligned}
    \right\},
    \quad \forall (x,y)\in\mathcal{D}
    \end{aligned}
\end{equation}
The lower-level optimizations consist of a chain of forecasting and solution maps $\mathcal{O}^i(\cdot)$, which take the previous decision as input. Similar to the machine learning task, each sample $(x,y)\in\mathcal{D}$ constructs an independent decision chain, whose outputs are eventually evaluated by the upper-level objective $\ell(\cdot)$.

Fig.~\ref{fig:dfl_general} views \eqref{eq:dfl_general} as neural forecaster with optimization layers. During training, the outcome of each block, including both trainable forecaster and unlearnable optimization layers, is passed as the input to the subsequent layer(s). The training loss function $\ell(\cdot)$, which may take the outcomes of all layers, must reflect the quality of the entire chain of decision-making. Notably, depending on the specific decision-making process \cite{xu2024e2e}, the number of forecasters and optimization models, as well as their types of interconnections can vary.

\subsection{DfL for Economic Dispatch and Redispatch Problems}

Although this paper presents a general efficient DfL framework as \eqref{eq:dfl_general}, we simulate its performance based on the chain of renewable/load forecast, economic dispatch (ED), and redispatch (RD). When the end-to-end decision chain is considered as in Fig.~\ref{fig:dfl_closedloop}, a \emph{closed-loop} DfL \cite{xu2025lapso} is defined from \eqref{eq:dfl_general}:
\begin{equation}\label{eq:closed_loop_dfl}
    \begin{aligned}
        \min_\theta ~~ & \textstyle \frac{1}{N_\mathcal{D}} \sum_{(x,y)\in\mathcal{D}} \ell(\hat{z}^{ed,\star}, \hat{z}^{rd,\star}) \\
    \text{s.t.} ~~ & \left\{
    \begin{aligned}
    & \hat{z}^{rd,\star} = \mathcal{O}^{rd}(\hat{z}^{ed,\star},y) \\
    & \hat{z}^{ed,\star} = \mathcal{O}^{ed}(\hat{y}), \; \hat{y} = f(x,\theta)
    \end{aligned}
    \right\}, \quad \forall (x,y)\in\mathcal{D}
    \end{aligned}
\end{equation}
where $\mathcal{O}^{ed}(\cdot)$ and $\mathcal{O}^{rd}(\cdot)$ represent the ED and RD optimizations, respectively. The training loss takes the decisions from both ED and RD stages to compute the system-wise operational cost \cite{xu2025lapso}.  In short, a closed-loop DfL setting trains the energy forecaster to minimize the expected end-to-end operational cost over the training dataset.

In contrast, it is possible to formulate DfL without end-to-end cost. Instead, an oracle label such as the optimal value of ED $f_0^{ed}(z^{ed,\star}(y))$ is computed from the true load and renewable $y$. As in real-time, the true energy forecast is unavailable at the ED stage, it is therefore referred to as \emph{counterfactual} DfL \cite{xu2025lapso}.
\begin{equation}\label{eq:counterfactual_dfl}
    \begin{aligned}
        \min_\theta ~~ & \textstyle \frac{1}{N_\mathcal{D}} \sum_{(x,y)\in\mathcal{D}} (f_0^{ed}(\hat{z}^{ed,\star}) - f_0^{ed}(z^{ed,\star}))^2 \\
    \text{s.t.} ~~ & \left\{
    \begin{aligned}
    & z^{ed,\star} = \mathcal{O}^{ed}(y) \\
    & \hat{z}^{ed,\star} = \mathcal{O}^{ed}(\hat{y}), \; \hat{y} = f(x,\theta)
    \end{aligned}
    \right\}, \quad \forall (x,y)\in\mathcal{D}
    \end{aligned}
\end{equation}
The oracle decisions $z^{ed,\star}(y)$ and their corresponding objective values can be computed and cached before training. The counterfactual DfL is illustrated in Fig.~\ref{fig:dfl_counterfactual}.

\section{Efficient Forward Pass}

\begin{listing}[t]%
\fontsize{9pt}{9pt}\selectfont
\caption{DfL Training Loop in Power Systems}%
\label{lst:dlf_training}%
\begin{lstlisting}[language=Haskell]
for epoch in range(N_epoch):
    for (X,Y) in batch_list:
        Forward NN: Y_hat = nn(X)
        Solve parallel: Z = O(Y_hat)
        loss = loss_func(Z,Y)
        loss.backward()
\end{lstlisting}
\end{listing}

\begin{figure}[t]
    \centering
    \includegraphics[width=0.7\linewidth,clip]{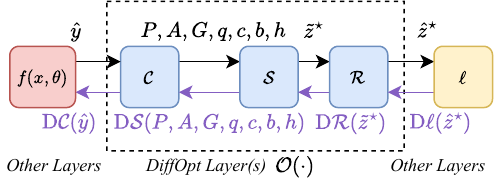}
    \caption{DiffOpt as part of NN. The forward and backward passes are in black and purple, respectively.}
    \label{fig:diffopt}
\end{figure}

A pseudo DfL training loop on Fig.~\ref{fig:dfl_general} is illustrated in Listing~\ref{lst:dlf_training}. To align with the SGD-based method in modern deep learning, the differentiation must be defined through the optimization layers $\mathcal{O}(\hat{y})$, which is referred to as \emph{differentiable optimization} (DiffOpt). Similar to CvxpyLayers, DiffAPQP designs DiffOpt functions with \emph{canonicalization}, \emph{solution}, and \emph{retrieval} steps (Fig.~\ref{fig:diffopt}). The canonicalization transforms \eqref{eq:ori_prob} into a standard problem format; the solution procedure calls a solver backend to solve the standard form problem; at last, the retrieval projects the optimal solution of standard problem into the original structure. Note that along the three steps, differentiation rules must be defined accordingly. This section considers designing efficient forward pass, which is highlighted by the black arrows in Fig.~\ref{fig:diffopt}. 

\subsection{Affine Parametric Quadratic Program}

Due to the computational efficiency and the guaranteed global optimum, many optimization problem in operation and market are formulated as quadratic program (QP). This includes direct current optimal power flow (OPF), ED, RD, market bidding, or even model predictive control (MPC), etc. The QP (or LP) formulation also dominates the DfL in power systems, whose scalability is challenging for training (see Table~\ref{tab:dfl_power_system_review}). This section starts by defining a class of structured QP parameterized by parameter $\hat{y}$.

\begin{definition}
    The QP $\mathcal{QP}(\hat{y})$ is affine-parametric if the linear objective term and right-hand-side (RHS) constraints depend affinely on parameters.
\end{definition}

Mathematically, APQP is denoted as,
\begin{equation}\label{eq:apqp}
    \begin{aligned}
        \min_{\tilde{z}} ~~ &  f_0(\tilde{z}) := \textstyle \frac{1}{2}\tilde{z}^TP\tilde{z} + q(\hat{y})^T\tilde{z} + c(\hat{y}) \\
        \text{s.t.} ~~ & A\tilde{z} = b(\hat{y}) \quad (\tilde{\nu}) \\
        & G\tilde{z} \leq h(\hat{y}) \quad (\tilde{\lambda})
    \end{aligned}
\end{equation}
In \eqref{eq:apqp}, $q(\hat{y}) := \tilde{q} + \sum_{i=1}^{N_p}Q_i\hat{y}_i$, $b(\hat{y}) := \tilde{b} + \sum_{i=1}^{N_p}B_i\hat{y}_i$, $h(\hat{y}) := \tilde{h} + \sum_{i=1}^{N_p} H_i\hat{y}_i$, and $c(\hat{y}) = \tilde{c} + \sum_{i=1}^{N_p}d_i^T\hat{y}_i$; $\tilde{z}\in\mathbb{R}^{n_d}$, $A\in\mathbb{R}^{n_{eq}\times n_d}$, and $G\in\mathbb{R}^{n_{in}\times n_d}$; $P\succeq 0$; $\tilde\nu\in\mathbb{R}^{n_{eq}}$ and $\tilde\lambda\in\mathbb{R}^{n_{in}}$ are the dual variables for the equality and inequality constraints. In addition, $\hat{y}_i$ retains the same definition as \eqref{eq:ori_prob} while $n_d$, $n_{eq}$, and $n_{in}$ denote the dimensions of the decision variable, equality constraints, and inequality constraints in APQP form.

\subsection{Canonicalization}\label{sec:canonicalization}

In \eqref{eq:apqp}, $(P,q,c,A,b,G,h)$ is considered as the \emph{data} of the QP while $\hat{y}$ is regarded as the problem \emph{parameter} as in \eqref{eq:ori_prob}. Many of the aforementioned operations can be reformulated to APQP. In the renewable forecast followed by the ED example, the forecast load and renewable will appear on the RHS of the constraints after rearrangement. Similarly, for the price forecast followed by the bidding example, the linear objective coefficient $q$ is an affine function of the parameter. An example of a simple PSO problem is illustrated as follows. Consider
\begin{equation}\label{eq:simple_pso}
    \begin{aligned}
        \min_{P_g} ~~ & c_g^TP_g \\
        \text{s.t.} ~~ & P_{bus} = C_gP_g - C_dP_d \\
        & -P_f^{max} \leq MP_{bus} \leq P_f^{max} \\
        & e^TP_{bus} = 0, ~~ P_g^{min} \leq P_g \leq P_g^{max},
    \end{aligned}
\end{equation}
where the decision variable is generator output $P_g$ and the parameter is the demand $P_{d}$. The data includes generator first-order cost $c_g$, incidence matrix $C_g$ and $C_d$, power transfer distribution factor $M$, transmission line thermal limit $P_f^{max}$, generator limit $P_g^{min}$ and $P_g^{max}$; $e$ is a vector whose entries are ones. As \eqref{eq:simple_pso} is defined by the physical meaning of each component, it is an example of \eqref{eq:ori_prob}, whose standard APQP form \eqref{eq:apqp} after canonicalization becomes,
\begin{equation*}
    \begin{array}{l}
        \tilde{z} = P_g, \quad \hat{y}_1 = P_d,\quad P = 0,\quad q=c_g, \quad \tilde{c} = 0, \quad d_1 = 0, \\
        G = \begin{pmatrix}
            MC_g \\
            -MC_g \\
            I \\
            -I
        \end{pmatrix}, \; h = \begin{pmatrix}
            P_f^{max} \\ P_f^{max} \\ P_g^{max} \\ -P_g^{min}
        \end{pmatrix}, \; H_1 = \begin{pmatrix}
            MC_d \\ -MC_d \\ 0 \\ 0
        \end{pmatrix}, \\
        A = e^TC_g, \quad b = 0, \quad B_1 = e^TC_d.
    \end{array}
\end{equation*}

As stated in Section~\ref{sec:sequential_chain}, it is essential to canonicalize \eqref{eq:ori_prob} into standard form before calling a particular solver. Obviously, when the PSO becomes complex, the manual canonicalization becomes time-consuming and prone to errors. We implement the automatic canonicalization step $(P,q(\hat{y}),c(\hat{y}),A,b(\hat{y}),G,h(\hat{y})) = \mathcal{C}(\hat{y})$ for \eqref{eq:ori_prob}, by tracking the relationships among parameters, data, and decision variables to avoid case-by-case manual derivation. Later in Section~\ref{sec:backward}, the benefit of this specific canonicalization is demonstrated when accelerating the backward pass differentiation. 

\subsection{Solution of APQP}
\label{sec:solution}

The optimal primal and dual pairs of APQP \eqref{eq:apqp} can be denoted as a solution map $(\tilde{z}^\star,\tilde{\lambda}^\star,\tilde{\nu}^\star) = \mathcal{S}(P,q,c,A,b,G,h)$. As $\mathcal{S}(\cdot)$ is an implicit representation, an iterative algorithm is required by calling a solver backend in the forward pass. Compared to CvxpyLayers, this paper investigates three solver-independent and broadly applicable strategies for accelerating iterative algorithms in the context of DfL for power system optimizations.

\subsubsection{Disentangled Forward and Backward Passes}\label{sec:disentable_forward_backward}

To start, in default setting, CvxpyLayers is built on top of DiffCP \cite{agrawal2019differentiating}, a Python package for differentiating \emph{conic} programs. Although conic program covers the quadratic program, this setting results in several limitations. 

First, conic program solvers can be inefficient to solve QPs, compared to dedicated QP solvers. Second, DiffCP integrates the forward and backward passes into a more compound algorithm where the solver backend is limited to SCS \cite{o2021operator}, ECOS \cite{domahidi2013ecos}, and Clarabel \cite{gaulart2024clarabel}. Although the original SCS and Clarabel solvers support quadratic objectives, DiffCP requires an \emph{epigraphical reformulation} into a linear objective. Consequently, slack variable $t$ is introduced so that the objective of \eqref{eq:apqp} becomes $\min_{\tilde{z}} t + q^T\tilde{z}$, with slack constraints $\tilde{z}^TP\tilde{z} \leq 2t$. Now assume $P$ is positive definite, the Cholesky factorization of $P$ is $P=F^TF$ where $F\in\mathbb{R}^{n_d\times n_d}$. Then the quadratic constraints can be replaced by a rotated cone, i.e., $(t,1,F\tilde{z}) \in \mathcal{Q}_r^{n_d+2}$. The $F\tilde{z}$ term in the cone introduces $n_d$ extra constraints and extra rotated cone of size $n_d+2$. These additional elements significantly increase problem size and computation time in the forward pass.

In contrast, the forward and backward passes in DiffAPQP is disentangled, meaning that the QPs can be solved by any suitable solver backends (Table~\ref{tab:compare}). This also suggests that DiffAPQP implements the original SCS and Clarabel which natively support quadratic objective without epigraphical reformulation \cite{o2021operator, gaulart2024clarabel}.

\subsubsection{Factorization Caching and Solver-data Update}\label{sec:solver_update}

Typically, solution algorithms for \eqref{eq:apqp} consist of a two-step procedure per iteration,
\begin{itemize}[leftmargin=*]
    \item \emph{Step One}. Formulate and solve a KKT system. 
    \item \emph{Step Two}. Apply the solution as an update to the primal and dual pairs. 
\end{itemize}
and the main computational burden is dominated by Step One where a linear system is solved.

Introduce slack variable $\zeta = Gz$ and scaled dual variable $u$. Using Alternating Direction
Method of Multipliers (ADMM) \cite{neal2011distributed} as an example, at iteration $k$, the following linear system needs to be solved for Step One:
\begin{equation}\label{eq:admm}
    \underbrace{\begin{pmatrix}
        P+\rho G^TG & A^T \\ A & 0
    \end{pmatrix}}_{K\in\mathbb{R}^{n_{d} + n_{eq}}}
    \begin{pmatrix}
        z^{k+1} \\ w^{k+1}
    \end{pmatrix} = 
    \begin{pmatrix}
        \rho G^T (\zeta^k - u^k) - q \\ b
    \end{pmatrix},
\end{equation}
where $w$ is the slack variable in equality-constrained $z$-step subproblem and $\rho$ is the step size \cite{neal2011distributed}. 

Now consider the DfL training loop in Listing~\ref{lst:dlf_training}. Although the training samples are distinct, the \emph{same} APQP template is repeatedly solved a total of $N_{epoch}\times N_{\mathcal{D}}$ times. As shown in \eqref{eq:apqp}, the affine structure guarantees that the KKT matrix $K$ of \eqref{eq:admm} remains \emph{unchanged} throughout the entire training process, given fixed $\rho$. Consequently, it is possible to factorize $K$ upon its first appearance for the first sample and reuse it for \emph{every} subsequent solves across \emph{all} samples and epochs. As the cubic flop factorization
dominates the computational burden in solving the linear system, factorization caching and reuse can offer significant speedup. 
Other solvers follows similar solution steps and Step One structure \eqref{eq:admm} so that the above factorization caching mechanism is generally applicable.

\subsubsection{Warm-Starting}

The effectiveness of warm-start strategy for repeated solves on similar problems has been demonstrated in many optimization algorithms such as OSQP \cite{stellato2020osqp}, SCS \cite{o2021operator}, and QPALM \cite{hermans2022qpalm}, etc. Similarly, when the DfL training approaches to convergence, the forecast variation for a given sample becomes small between epochs and warm-start strategies become effective. In detail, given one training sample $(x,y) \in \mathcal{D}$, let the sequence of same forecasts across all epochs be denoted by $\mathcal{Y} = \{\hat{y}^1, \ldots, \hat{y}^{N_{epoch}}\}$. The initialization at epoch $j+1$ can reuse the solution from the previous run, $\mathcal{S}(\hat{y}^j)$, including the solver-native primal and dual variables as well as adapted parameters.  

To sum up, the DiffAPQP accelerates the forward pass against CvxpyLayers from three aspects: (i). the ``disentangled forward and backward passes'' represent a universal property of DiffAPQP for different applications and solver choices; (ii). ``solver-data update'' is a training sample independent acceleration; while (iii). ``warm-starting'' is sample dependent acceleration.

\subsection{Retrieval from APQP}

When the optimal solution $\tilde{z}^\star$ of \eqref{eq:apqp} is available from solution step, a retrieval step is designed to obtain the solution of the original problem \eqref{eq:ori_prob}, e.g., $\hat{z}^\star = \mathcal{R}(\tilde{z}^\star)$. Based on \eqref{eq:ori_prob} and the affine structure of \eqref{eq:apqp}, $\mathcal{R}(\cdot)$ is a linear map which consists of block permutation matrix, depending on the sequence of $\hat{z}_i,\forall i$ in $\tilde{z}$ during canonicalization. Let $\hat{z}_i^\star$ be the $i$-th optimal decision variable, then 
\begin{equation}\label{eq:retrieval}
    \hat{z}^\star_i = \mathcal{R}(\tilde{z}^\star) = \begin{pmatrix}
        0 & \cdots & I & \cdots & 0
    \end{pmatrix}\tilde{z}^\star.
\end{equation}

For DfL, $\hat{z}^\star$ is fed into a scalar loss function $l=\ell(\hat{z}^\star)$ in \eqref{eq:dfl_general}, to evaluate the quality of the decision-making process.  To sum up, the forward pass for DiffOpt can be denoted as a function composition \eqref{eq:forward_pass} (see Fig.~\ref{fig:diffopt}). Note that $\ell(\cdot)$ can contain multiple layers of neural networks or other DiffOpt layers to represent a complex decision-making chain (see Fig.~\ref{fig:dfl}). However, since the final training loss is always scalar, we omit the complex dependence for simplicity purposes.
\begin{equation}\label{eq:forward_pass}
    l = \ell \circ \mathcal{O} \circ f = \ell \circ \mathcal{R} \circ \mathcal{S} \circ \mathcal{C} \circ f.
\end{equation}

\section{Efficient Backward Pass}\label{sec:backward}

\subsection{Differentiation on Full KKT System of Solution Map}

\subsubsection{Derivation}

In the previous section, the optimal primal and dual variables are solved via the efficient forward pass. In the backward pass, it is necessary to differentiate through the solver $\mathcal{O}(\cdot) = \mathcal{R}\circ\mathcal{S}\circ\mathcal{C}$,
\begin{subequations}\label{eq:backward_pass}
\begin{equation}\label{eq:backward_pass_entire}
    \mathrm{D}_\theta\ell = \mathrm{D} \ell(\hat{z}^\star)\cdot \mathrm{D} \mathcal{O}(\hat{y}) \cdot \mathrm{D}_\theta f(x,\theta), \quad \text{with}
\end{equation}
\begin{equation}\label{eq:backward_pass_optimization}
    \mathrm{D} \mathcal{O}(\hat{y}) = \mathrm{D}\mathcal{R}(\tilde{z}^\star)\cdot \mathrm{D}\mathcal{S}(P,q,c,A,b,G,h) \cdot \mathrm{D} \mathcal{C}(\hat{y}).
\end{equation}
\end{subequations}
In \eqref{eq:backward_pass_entire}, $\mathrm{D} \ell(\hat{z}^\star)$ and $\mathrm{D}_\theta f(x,\theta)$ can be obtained by automatic differentiation (AD) such as PyTorch \cite{paszke2019pytorch}. The main computational difficulty remains in differentiating the optimization problem. The differentiation starts from left to right in the reverse mode automatic differentiation (RMAD) \cite{blondel2024elements}, meaning that the outputs of all layers, including both NNs and DiffOpts, are stored in memory, and their relationship must be recorded on the computational graph in the forward pass.

First, it is not difficult to see from \eqref{eq:retrieval} that $\mathrm{D}\mathcal{R}(\tilde{z})$ is a linear block permutation matrix. Second, instead of differentiating through the unrolled solver $\mathcal{S}(\cdot)$ such as \cite{kotary2023backpropagation}, the implicit function theorem \cite{blondel2024elements} can be used to differentiate the equality-constrained part of the KKT condition \eqref{eq:kkt_full}, which is compactly denoted as $\mathcal{K}(\tilde{z}^\star, \tilde{\lambda}^\star, \tilde{\nu}^\star, \hat{y}) = 0$. Considering the affine structure of $b$, $h$, and $q$ on $\hat{y}_i$s in \eqref{eq:apqp}, the \emph{full} differentiation of $\mathcal{K}(\cdot)$ with respect to the $i$-th parameter $\hat{y}_i$ is derived as,
\begin{subequations}\label{eq:diff_kkt_full}
    \begin{equation}
        K_{a}\cdot \mathrm{D}_{\hat{y}_i}(\tilde{z}^\star, \tilde{\nu}^\star, \tilde{\lambda}^\star) = D_{a,i},\quad \text{with}
    \end{equation}
    \begin{equation}
        K_{a} = \begin{pmatrix}
        P & A^T & G^T \\
        A & 0 & 0 \\
        \dg{(\tilde{\lambda}^\star)}G & 0 & \dg{(G\tilde{z}^\star-h) }
    \end{pmatrix},
    \end{equation}
    \begin{equation}
        \mathrm{D}_{\hat{y}_i}(\tilde{z}^\star, \tilde{\nu}^\star, \tilde{\lambda}^\star) = \begin{pmatrix}
        \mathrm{D}_{\hat{y}_i} \tilde{z}^\star \\
        \mathrm{D}_{\hat{y}_i} \tilde{\nu}^\star \\
        \mathrm{D}_{\hat{y}_i} \tilde{\lambda}^\star
    \end{pmatrix}, D_{a,i} = \begin{pmatrix}
        -Q_i \\
        B_i \\
        \dg{(\tilde{\lambda}^\star)}H_i
    \end{pmatrix}.
    \end{equation}
\end{subequations}
where $\dg{(\cdot)}$ is the diagonal operator. Note that $K_{a}$ is not symmetric and \eqref{eq:diff_kkt_full} already obtains the differentiation with respect to the original parameter $\hat{y}_i$. In addition, thanks to the canonicalization of APQP \eqref{eq:apqp}, \eqref{eq:diff_kkt_full} is analytical and no AD is needed. Consequently, the affine canonicalization maps are evaluated analytically through
$D_{a,i}$ and \eqref{eq:backward_pass} becomes,
\begin{equation}\label{eq:backward_pass_1}
    \textstyle \mathrm{D}_\theta\ell = \underbrace{\mathrm{D}\ell(\hat{z}^\star)\cdot\mathrm{D}\mathcal{R}(\tilde{z}^\star) \cdot K_a^\dagger}_{=s_a^T\in\mathbb{R}^{1\times(n_d+n_{in}+n_{eq})}} \cdot \sum_{i=1}^{N_p} \left(D_{a,i}\cdot D_\theta f_i(x,\theta)\right),
\end{equation}
where $f_i(x,\theta)$ represents the forecaster output corresponding to the $i$-th forecast parameter.
The pseudo-inverse is used in case $K_a$ is not invertible. When RMAD is applied, it is not necessary to explicitly take the pseudo-inverse. Considering \eqref{eq:backward_pass_1}, the row vector $s_a^T$ can be viewed as the adjoint of $\tilde{z}^\star$ and the following least-square problem is solved,
\begin{equation}\label{eq:least_square}
    s^\star_{a} = \arg\min_{s_a} \|K_{a}^T\cdot s_a - \mathrm{D}^T\mathcal{R}(\tilde{z}^\star)\cdot\mathrm{D}^T\ell(\hat{z}^\star)\|^2.
\end{equation}
Since $K_{a}$ is asymmetric and likely sparse, the LSQR \cite{paige1982lsqr} is used to iteratively solve \eqref{eq:least_square} as in CvxpyLayers. Plugging $s_a^\star$ into \eqref{eq:backward_pass_1}, the full differentiation on \eqref{eq:backward_pass} becomes
\begin{equation}\label{eq:backward_pass_2}
    \textstyle \mathrm{D}_\theta\ell = s^{\star,T}_a \cdot\sum_{i=1}^{N_p} \left(D_{a,i} \cdot \mathrm{D}_\theta f_i(x,\theta)\right).
\end{equation}
Taking the transpose of \eqref{eq:backward_pass_2}, the computation of the gradient only involves matrix-vector products, which avoids computationally complex matrix-by-matrix products and explicit matrix inversion in \eqref{eq:backward_pass_1}. 

\subsubsection{Discussion}

For the APQP \eqref{eq:apqp}, the dimension of the linear system \eqref{eq:diff_kkt_full} is $(n_d+n_{eq}+n_{in})$ which is as large as the optimization itself so that the differentiation method is referred to as full-KKT differentiation. Recall the worst computational complexity is up to cubic of the system size; differentiation can account significant training time. Moreover, when the optimal solution violates the regularity conditions such as linear independence constraint qualification (LICQ), the KKT system becomes singular \cite{nocedal1999numerical}
so that the solution-map derivative may be non-unique or numerically unstable. 


\subsection{Differentiation on Reduced-order KKT System of Solution Map}\label{sec:reduced_kkt}

Since the optimal primal and dual pair are available after the forward pass, it is possible to use this information to reduce the size of $K_{a}$ in \eqref{eq:diff_kkt_full}. To start, let $\mathcal{A}=\{j\in\{1,\cdots,n_{in} \}|G_j\tilde{z}^\star = h_j \}$ be the index set of active inequality constraints. Moreover, let $(\cdot)_\mathcal{A}$ be the rows indexed by $\mathcal{A}$ of a matrix or vector. Similar definitions are made on the inactive inequality constraint set $\mathcal{I}$ and $(\cdot)_{\mathcal{I}}$. After the active inequality constraints are identified, \eqref{eq:apqp} is reduced into an equality-constrained QP \cite{boyd2004convex}:
\begin{equation}\label{eq:apqp_reduced}
    \begin{aligned}
        \min_{\tilde{z}} ~~ &  f_0(\tilde{z}) := \textstyle \frac{1}{2}\tilde{z}^TP\tilde{z} + q(\hat{y})^T\tilde{z} \\
        \text{s.t.} ~~ & \textstyle A\tilde{z} = \tilde{b} + \sum_{i=1}^{N_p}B_i\hat{y}_i \quad (\tilde{\nu}) \\
        & \textstyle G_\mathcal{A}\tilde{z} = \tilde{h}_{\mathcal{A}} + \sum_{i=1}^{N_p} H_{i,\mathcal{A}}\hat{y}_i \quad (\tilde{\lambda}_\mathcal{A})
    \end{aligned}
\end{equation}
The differentiation on the KKT system of \eqref{eq:apqp_reduced} becomes,
\begin{subequations}\label{eq:diff_kkt_reduced}
    \begin{equation}
        K_{s}\cdot\mathrm{D}_{\hat{y}_i}(\tilde{z}^\star, \tilde{\nu}^\star, \tilde{\lambda}_{\mathcal{A}}) = D_{s,i},\quad \text{with}
    \end{equation}
    \begin{equation}
        K_{s} = \begin{pmatrix}
        P & A^T & G_{\mathcal{A}}^T \\
        A & 0 & 0 \\
        G_{\mathcal{A}} & 0 & 0
    \end{pmatrix}, \quad D_{s,i} = \begin{pmatrix}
        -Q_i \\
        B_i \\
        H_{i,\mathcal{A}}
    \end{pmatrix}.
    \end{equation}
\end{subequations}
Notably $K_s$ is symmetric such that \eqref{eq:least_square} becomes,
\begin{equation}\label{eq:minres}
    s^\star_{s} = \arg\min_{s_s} \|K_{s}\cdot s_s - \mathrm{D}^T\mathcal{R}(\tilde{z}^\star)\cdot\mathrm{D}^T\ell(\hat{z}^\star)\|^2,
\end{equation}
and the full differentiation in \eqref{eq:backward_pass_2} becomes
\begin{equation}\label{eq:backward_pass_3}
    \textstyle \mathrm{D}_\theta\ell = s^{\star,T}_s \cdot\sum_{i=1}^{N_p} \left(D_{s,i} \cdot \mathrm{D}_\theta f_i(x,\theta)\right).
\end{equation}

The advantages of canonicalizing \eqref{eq:ori_prob} into APQP \eqref{eq:apqp} become evident as both $K_{s}$ and $D_{s}$ depend solely on the problem data, rather than on the primal and dual variables. The full and reduced dimension are $n_d+n_{eq}+n_{in}$ and $n_{d}+n_{eq}+|\mathcal{A}|$, respectively; the benefit of reduced system \eqref{eq:diff_kkt_reduced} grows when $|\mathcal{A}| \ll n_{in}$. The power system optimizations likely follow this condition due to the redundant constraints on the generator capacities and transmission line thermal limits, etc. More importantly, as $K_{s}$ is symmetric, more efficient minimum residual iteration (MINRES) \cite{paige1975solution} can be used to solve the least squares system \eqref{eq:minres}, compared to LSQR for the asymmetric case in CvxpyLayers and full-KKT case.

Moreover, the full KKT matrix $K_a$ in \eqref{eq:diff_kkt_full} can become ill-conditioned because its complementarity blocks contain inequality slacks and dual multipliers close to zero. Consequently, solving the adjoint system in \eqref{eq:least_square} becomes numerically unstable. The reduced-KKT formulation can improve the robustness of the adjoint computation by retaining only the numerically active inequalities. Specifically, the active set is identified as $\mathcal{A}=\{j\in\{1,\cdots,n_{in}\} \mid |G_j\tilde{z}^\star - h_j| \leq \max\{\epsilon_{abs}, \epsilon_{rel}(1+|h_j|) \}$. In practice, $\varepsilon_{abs}$ should be sufficiently larger than the feasibility tolerance of the forward pass optimization solver to avoid excluding constraints that are binding up to numerical error. However, an excessively large tolerance may include nearly active or linearly dependent constraints, making the reduced KKT system singular or poorly conditioned. Thus, improved robustness is obtained only when the active set is identified reliably and satisfies the required regularity conditions. 



\subsection{Adjoint-free Value-function Differentiation}


Consider the APQP \eqref{eq:apqp} with value function $\alpha(\hat{y})$. The counterfactual DfL training in Fig.~\ref{fig:dfl_counterfactual} directly takes the value function as input to the loss function $\ell(\alpha(\hat{y}))$. As an optimal primal and dual pair $(\tilde{z}^\star, \tilde{\lambda}^\star, \tilde{\nu}^\star)$ is obtained during the forward pass and based on the perturbation analysis \cite{boyd2004convex}, the differentiation of the value function with respect to the parameter can be derived as
\begin{equation}\label{eq:pso_differentiation_free}
    \begin{aligned}
        \mathrm{D}_{\hat{y}_i}\alpha(\hat{y}) & = \mathrm{D}_{q}\alpha(\hat{y})\cdot\mathrm{D}_{\hat{y}_i}q + \mathrm{D}_c\alpha(\hat{y})\cdot\mathrm{D}_{\hat{y}_i}c  \\
        & ~~~ + \mathrm{D}_{b}\alpha(\hat{y})\cdot \mathrm{D}_{\hat{y}_i}b + \mathrm{D}_{h}\alpha(\hat{y})\cdot \mathrm{D}_{\hat{y}_i}h \\
        & = \tilde{z}^{\star,T}Q_i + d_i^T -\tilde{\nu}^{\star,T}B_i - \tilde{\lambda}^{\star,T}H_i.
    \end{aligned}
\end{equation}
Consequently, the full differentiation \eqref{eq:backward_pass} is derived from chain rule,
\begin{equation}\label{eq:backward_pass_free}
    \textstyle \mathrm{D}_\theta\ell = \mathrm{D}_\alpha\ell(\alpha(\hat{y}))\cdot\sum_{i=1}^{N_p} \left(\mathrm{D}_{\hat{y}_i}\alpha(\hat{y}) \cdot \mathrm{D}_\theta f_i(x,\theta)\right).
\end{equation}
Compared to differentiation on the solution map,
all the items in \eqref{eq:backward_pass_free} are available during the forward pass. I.e., no extra implicit differentiation step or solving linear system is needed as in full-KKT \eqref{eq:least_square} or reduced-KKT \eqref{eq:minres} cases. Therefore, the value function map avoids the singularity difficulty. However, it does not indicate the direct differentiation on the optimal value function results in unique gradient.

\subsection{Theoretical Results}\label{sec:theoretical_result}

In this section, we demonstrate the theoretical results for the efficient differentiation method presented in Section~\ref{sec:backward}.

For certain irregular cases, it is possible to have singular $K_a$ and $K_s$ such that the outcomes of \eqref{eq:least_square} and \eqref{eq:minres} are sub-gradients. We therefore prove a sufficient condition to have non-singular $K_{a}$ and $K_{s}$, based on which, the equivalence between full KKT system \eqref{eq:diff_kkt_full} and reduced KKT system \eqref{eq:diff_kkt_reduced} are proved. To start, consider the following assumptions,
\begin{itemize}[leftmargin=*]
        \item (A1-KKT Condition). The forward pass solution $(\tilde{z}^\star, \tilde{\lambda}^\star, \tilde{\nu}^\star)$ satisfies the KKT conditions,
    \begin{equation}\label{eq:kkt_full}
        \begin{aligned}
         P \tilde{z}^{\star}+q+A^T \tilde{\nu}^{\star}+G^T \tilde{\lambda}^{\star} & =0, ~~ A \tilde{z}^{\star}-b =0 \\
        G \tilde{z}^\star -h & \leq 0, ~~ \tilde{\lambda}^\star \geq 0 \\
        \dg{(\tilde{\lambda}^{\star})} \cdot \left(G \tilde{z}^{\star}-h\right) & =0
    \end{aligned}
    \end{equation} 
    \item (A2). $\tilde{\lambda}^\star$ satisfies the strict complementarity condition at $\tilde{z}^\star$. E.g., $\tilde{\lambda}^\star_{\mathcal{A}} > 0$. 
    \item (A3). $P$ is positive definite on the null space of $(A^T, G_{\mathcal{A}}^T)^T$.
    \item (A4-LICQ). The gradients of the active constraints are linearly independent, e.g., $(A^T,G_\mathcal{A}^T)^T$ is full row rank. 
\end{itemize}
Under LICQ (A4), (A1) states the first-order KKT conditions. Assumption (A3) is the second-order sufficient condition on the active-constraint null space, while (A2) imposes strict complementarity \cite{nocedal1999numerical}. 

\begin{proposition}\label{prop:kkt_equivalence}
    Let $(\mathrm{D}\tilde{z}^{a}, \mathrm{D}\tilde{\lambda}^{a}, \mathrm{D}\tilde{\nu}^{a})$ and $(\mathrm{D}\tilde{z}^{s}, \mathrm{D}\tilde{\lambda}^{s}_{\mathcal{A}}, \mathrm{D}\tilde{\nu}^{s})$ be the solutions to \eqref{eq:diff_kkt_full} and \eqref{eq:diff_kkt_reduced}, respectively. If (A1)-(A4) are satisfied, then (i). $K_{a}$ and $K_{s}$ are non-singular; (ii). $\mathrm{D}\tilde{\lambda}_{\mathcal{I}}^{a} = 0$; and (iii). $(\mathrm{D}\tilde{z}^{a}, \mathrm{D}\tilde{\lambda}^{a}_{\mathcal{A}}, \mathrm{D}\nu^{a}) = (\mathrm{D}\tilde{z}^{s}, \mathrm{D}\tilde{\lambda}^{s}_{\mathcal{A}}, \mathrm{D}\tilde{\nu}^{s})$.
\end{proposition}

Next, consider the linear objective (APLP) version of \eqref{eq:apqp}. Since $P=0$, (A3) is no longer satisfied. However, the main conclusions still hold under the following result.

\begin{proposition}\label{prop:kkt_equivalence_lp}
    Consider APLP with $P=0$. Take the same settings as in Proposition~\ref{prop:kkt_equivalence} and assume \textbf{(A1)}, \textbf{(A2)}, and \textbf{(A5)} a non-degenerate vertex solution is found in the solution step. Then all the conclusions in Proposition~\ref{prop:kkt_equivalence} are satisfied.
\end{proposition}

Recall that \eqref{eq:diff_kkt_full} and \eqref{eq:diff_kkt_reduced} may encounter singular issues as shown in previous research \cite{wilder2019melding,paredes2025participation}. Intuitively, the differentiation-free method \eqref{eq:backward_pass_free} should also have the similar problem. This is caused by the non-differentiability of $\alpha(\hat{y})$ at certain $\hat{y}$. It is also the case when there exist multiple dual variables, although the optimization solver only outputs one of them in the forward pass. In this case, it can be shown that \eqref{eq:pso_differentiation_free} becomes a sub-gradient for any given optimal dual variable as shown by Proposition~\ref{prop:subgradient_qbh}

\begin{proposition}\label{prop:subgradient_qbh} 
Let $\alpha(q,c,b,h)$ be the value function of \eqref{eq:apqp} with respect to the program data $q$, $c$, $b$, and $h$, i.e., 
\begin{equation} 
    \textstyle \alpha(q,c,b,h) := \inf_{\tilde z} \left\{ \frac{1}{2}\tilde z^{T}P\tilde z +q^{T}\tilde z+c \;\middle|\; A\tilde z=b,\; G\tilde z\leq h \right\}
\end{equation}   

Assume that Slater's condition holds for every considered $(b,h)$. Define the optimal primal solution set as $ Z^{\star}(q,b,h) := \{ \tilde z \; |\; (\tilde z,\tilde\nu,\tilde\lambda) \text{ satisfies the KKT conditions for } \eqref{eq:apqp} \text{ for some } (\tilde\nu,\tilde\lambda) \}, $ and the optimal dual solution set as $ \Lambda^{\star}(q,b,h) := \{ (\tilde\nu,\tilde\lambda) \; | \; (\tilde z,\tilde\nu,\tilde\lambda) \text{ satisfies the KKT conditions for } \eqref{eq:apqp} \text{ for some }\tilde z \}$, which are nonempty and bounded. Then: (i). $\alpha(q,c,b,h)$ is concave in $q$, affine in $c$, and convex in $(b,h)$; (ii). $ \mathrm{D}_{c}\alpha(q,c,b,h)=1$; (iii). $ \partial_{q}^{+}\alpha(q,c,b,h) = Z^{\star}(q,b,h), $ where $\partial_{q}^{+}$ denotes the superdifferential with respect to $q$; (iv). $ \partial_{(b,h)}\alpha(q,c,b,h) = \{ (-\tilde\nu,-\tilde\lambda) \;|\; (\tilde\nu,\tilde\lambda) \in\Lambda^{\star}(q,b,h)\} $; (v). if $ Z^{\star}(q,b,h)=\{\tilde z^{\star}\}, $ then $\alpha$ is differentiable with respect to $q$ and $ \mathrm{D}_{q}\alpha(q,c,b,h) = \tilde z^{\star,T} $; (vi). if $ \Lambda^{\star}(q,b,h) = \{(\tilde\nu^{\star},\tilde\lambda^{\star})\}, $ then $\alpha$ is differentiable with respect to $(b,h)$ and $ \mathrm{D}_{b}\alpha(q,c,b,h) = -\tilde\nu^{\star,T}$, $\mathrm{D}_{h}\alpha(q,c,b,h) = -\tilde\lambda^{\star,T}$.
\end{proposition}

Proposition~\ref{prop:subgradient_qbh} demonstrates that the differentiation \eqref{eq:backward_pass_free} follows the same ``singularity'' issue as \eqref{eq:diff_kkt_full} and \eqref{eq:diff_kkt_reduced} due to the multiple optimal primal and dual solutions. However, it will not introduce any computational inconvenience when solving linear system \eqref{eq:least_square} or \eqref{eq:minres}.

\section{Simulations}\label{sec:simulation}

\subsection{Simulation Settings}

The performance of the proposed DiffAPQP framework is compared with CvxpyLayers on the IEEE 118-bus system under closed-loop and counterfactual DfL settings, as shown in Fig.~\ref{fig:dfl}. The test system contains 34 generators, 10 solar plants, and 10 wind plants. Standard ED and RD models are formulated into continuous convex quadratic programs with 24-hour optimization horizon. 
This problem size provides a large-scale stress test for differentiable optimization in end-to-end DfL, which is not seen in the literature. One year of hourly nodal load, solar, wind, and weather data is obtained from the quasi-realistic Texas system~\cite{lu2025synthetic} as the forecasting target. The IEEE 118-bus testbed, profile assignment, and data preprocessing are implemented using our open-source GridForge package~\cite{xu2025lapso}. 

We employ a shared multi-task neural forecaster to jointly predict the 24-hour trajectories of system loads, solar generation, and wind generation. The model consists of three components: a normalized spatiotemporal input representation, a shared two-dimensional convolutional encoder, and three asset-specific forecasting heads. To start, AbL-based forecaster is trained with MSE loss defined in Section~\ref{sec:sequential_chain}, followed by DfL fine-tuning \eqref{eq:closed_loop_dfl} or \eqref{eq:counterfactual_dfl}. 
In DiffAPQP, warm starts are indexed using each sample's unique start-hour identifier, allowing a daily problem to retrieve its previous solution across epochs. During the DfL training, five solver backends, including operator splitting-based SCS and OSQP, interior point-based Clarabel and Gurobi, as well as augmented Lagrangian-based QPALM are tested with i) no warm-start and update (referred to as cold-start, CS); ii) with warm-start only (WS); iii) with update only (UP), and iii) with both warm-start and update (WU). All the experiments are tested on Ubuntu 20.04.6 LTS with two Intel(R) Xeon(R) Gold 6230R CPUs and \emph{with the number of CPUs capped at 32} to simulate a stress test environment.
All DiffAPQP and CvxpyLayers experiments were run under the same random seed, data split, model architecture, optimization solver arguments (for the same solver backend), batch size, and training configuration to ensure a fair comparison.

\begin{figure*}
     \centering
     \begin{subfigure}[b]{0.49\textwidth}
         \centering
         \includegraphics[width=\textwidth]{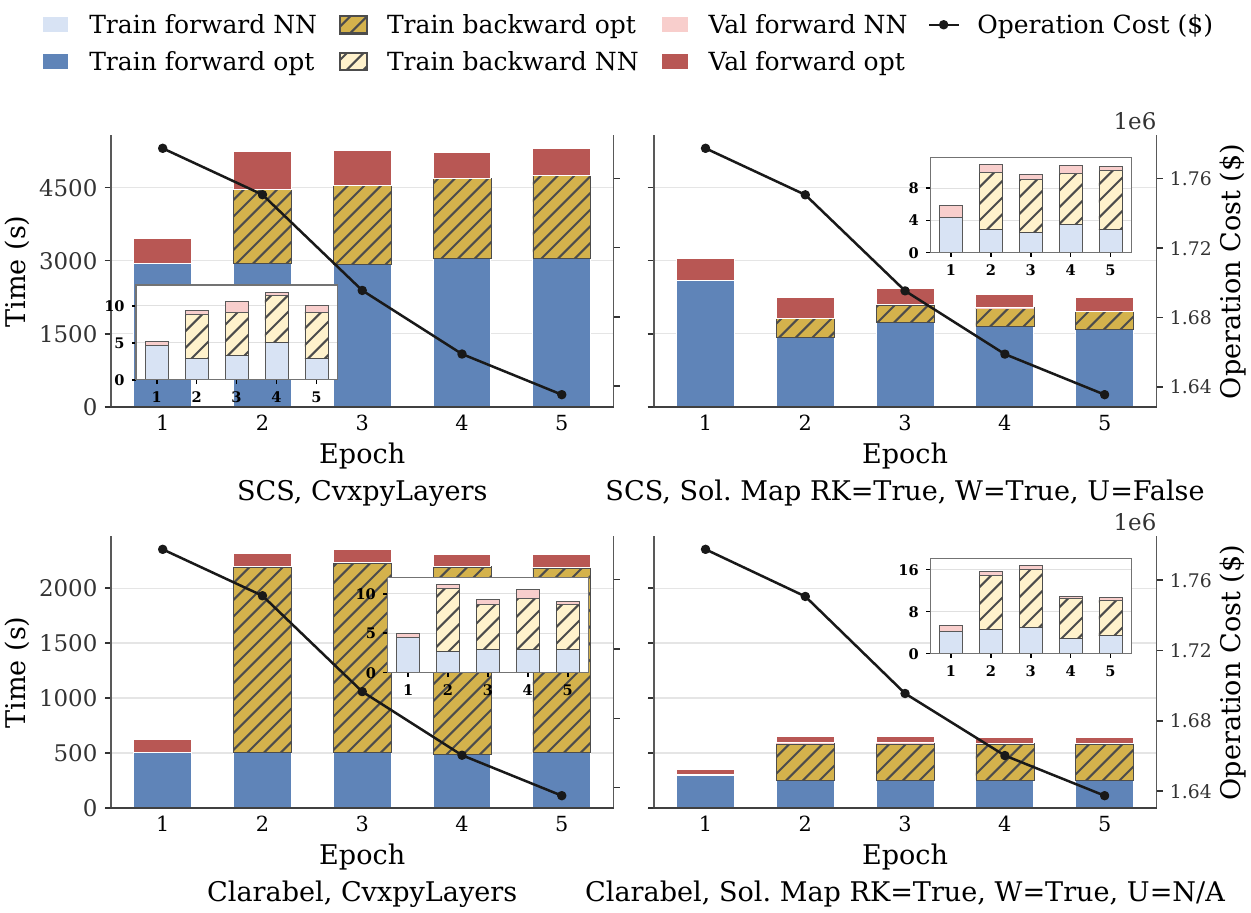}
         \caption{Closed-loop DfL}
         \label{fig:sim_endtoend_closed}
     \end{subfigure}
     \hfill
     \begin{subfigure}[b]{0.49\textwidth}
         \centering
         \includegraphics[width=\textwidth]{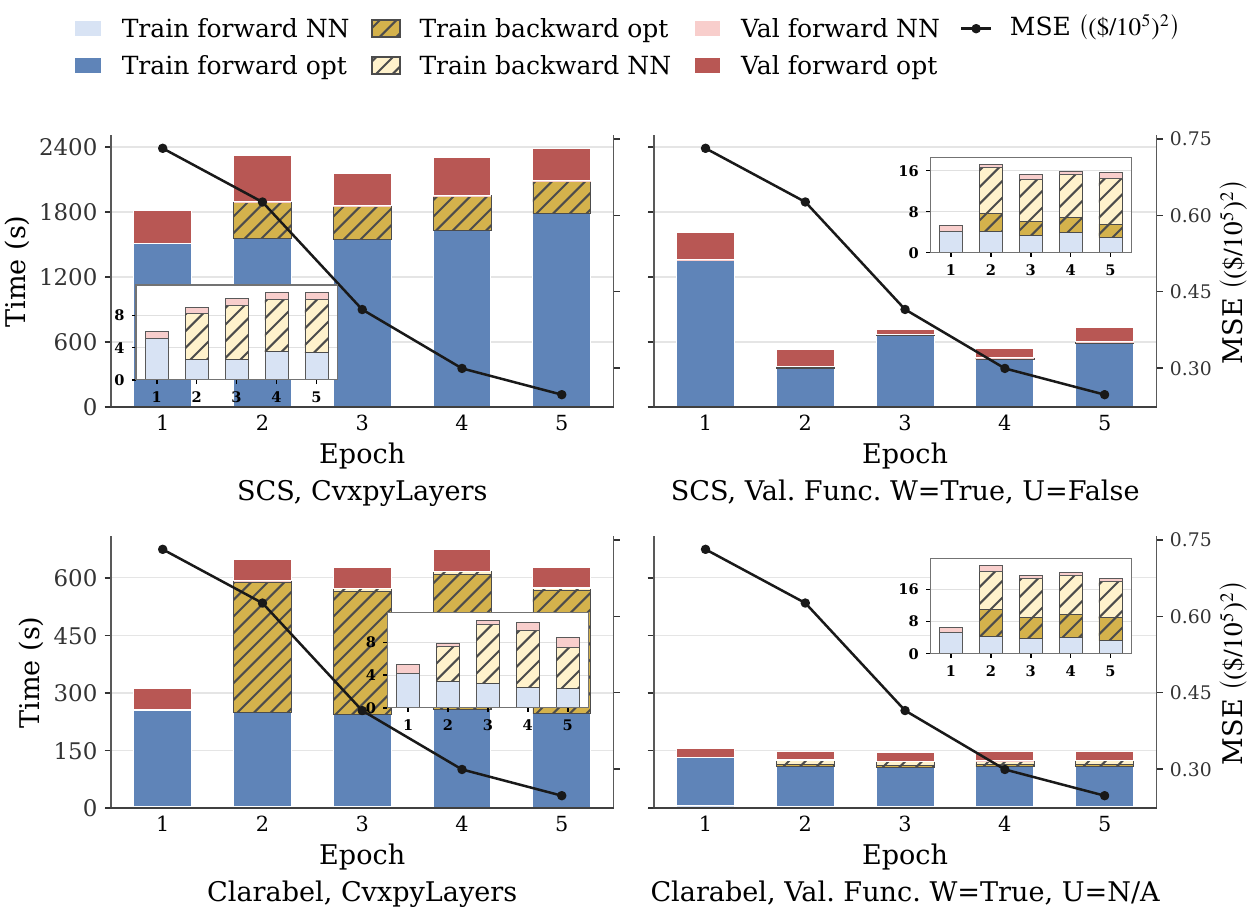}
         \caption{Counterfactual DfL}
         \label{fig:sim_endtoend_counter}
     \end{subfigure}
        \caption{End-to-end DfL training performances of DiffAPQP compared with CvxpyLayers using SCS and Clarabel solver backends.}
        \label{fig:sim_endtoend}
\end{figure*}

\begin{table*}[t]
\centering
\caption{Closed-loop DfL performance for all evaluated solver configurations. The overall epoch time includes
the recorded training forward/backward and validation forward components.}
\label{tab:closed_loop_resource_performance}
    \small
    \begin{adjustbox}{width=\textwidth}
    \begin{tabular}{
        l l c c c
        S[table-format=2.2]
        S[table-format=2.2]
        S[table-format=2.2]
        S[table-format=2.2]
        S[table-format=2.2]
        S[table-format=7.0]
    }
    \toprule
    \textbf{Method}
    & \textbf{Solver}
    & \textbf{RK}
    & \textbf{W}
    & \textbf{U}
    & {\textbf{Avg. cores}}
    & {\textbf{Peak memory (GiB)}}
    & {\textbf{Train opt. fwd. (min)}}
    & {\textbf{Train opt. bwd. (min)}}
    & {\textbf{Avg. epoch time (min)}}
    & {\textbf{Test cost (\$)}} \\
    \midrule
    
    CvxpyLayers & SCS & -- & -- & -- &
    14.65 & 59.53 & 49.81 & 26.88 & 87.43 & 1955769 \\
    DiffAPQP & SCS & True & False & False &
    17.75 & 26.55 & 43.54 & 6.82 & 58.57 & 1958673 \\
    DiffAPQP & SCS & True & False & True &
    17.66 & 26.77 & 45.73 & 6.17 & 60.68 & 1959441 \\
    DiffAPQP & SCS & True & True & False &
    14.08 & 26.74 & 26.63 & 6.21 & 38.44 & 1956677 \\
    DiffAPQP & SCS & True & True & True &
    13.81 & 26.78 & 35.22 & 6.06 & 47.65 & 1961202 \\
    
    \addlinespace
    CvxpyLayers & Clarabel & -- & -- & -- &
    17.55 & 56.46 & 8.32 & 28.27 & 38.65 & 1957534 \\
    DiffAPQP & Clarabel & False & True & {N/A} &
    23.80 & 28.51 & 4.22 & 12.82 & 18.22 & 1959320 \\
    DiffAPQP & Clarabel & True & False & {N/A} &
    21.74 & 28.50 & 5.30 & 5.82 & 12.53 & 1961295 \\
    DiffAPQP & Clarabel & True & True & {N/A} &
    20.73 & 28.56 & 4.17 & 5.47 & 10.79 & 1961295 \\
    
    \addlinespace
    DiffAPQP & OSQP & True & False & False &
    19.43 & 28.23 & 22.30 & 6.69 & 33.57 & 1963473 \\
    DiffAPQP & OSQP & True & False & True &
    17.57 & 28.23 & 16.01 & 7.13 & 26.61 & 1961587 \\
    DiffAPQP & OSQP & True & True & False &
    17.46 & 28.25 & 12.59 & 6.50 & 22.04 & 1953440 \\
    DiffAPQP & OSQP & True & True & True &
    15.68 & 28.16 & 10.44 & 6.31 & 18.96 & 1951734 \\
    
    \addlinespace
    DiffAPQP & QPALM & True & False & False &
    22.98 & 28.76 & 9.81 & 5.92 & 18.08 & 1956886 \\
    DiffAPQP & QPALM & True & False & True &
    22.20 & 28.77 & 9.41 & 5.94 & 17.66 & 1956886 \\
    DiffAPQP & QPALM & True & True & False &
    17.54 & 28.74 & 4.16 & 5.45 & 10.92 & 1960899 \\
    DiffAPQP & QPALM & True & True & True &
    20.96 & 28.73 & 7.98 & 5.79 & 16.00 & 1961098 \\
    
    \addlinespace
    DiffAPQP & Gurobi & True & False & {N/A} &
    13.54 & 27.23 & 3.29 & 5.72 & 9.88 & 1961920 \\
    DiffAPQP & Gurobi & True & True & {N/A} &
    12.28 & 27.25 & 3.94 & 5.65 & 10.61 & 1961920 \\
    
    \bottomrule
    \end{tabular}
    \end{adjustbox}
\end{table*}

\subsection{Performance on Computational Time}

As the evaluated CvxpyLayers 1.0.0 DiffCP interface does not expose OSQP, QPALM, or Gurobi as forward backends, to have a fair comparison, Figs.~\ref{fig:sim_endtoend_closed} and~\ref{fig:sim_endtoend_counter} compare the epoch-wise computational cost and learning trajectories of CvxpyLayers and DiffAPQP using SCS and Clarabel backends. The closed-loop and counterfactual DfLs implement the solution-map and value function layer, respectively. In all experiments, the first recorded epoch evaluates the pretrained AbL forecaster without backward pass; gradient-based DfL fine-tuning is performed during the remaining four epochs and all quantitative runtime comparisons are based on the average of epochs 2--5. First, across both formulations, the optimization layer dominates the total epoch time, whereas the NN forward and backward passes account for only a small fraction of the runtime, as highlighted by the insets. 

In the closed-loop formulation, the gradient of the realized operating cost must propagate through the coupled RD and ED solution maps. For SCS, the selected reduced-KKT DiffAPQP in WS mode results in a $2.27\times$ end-to-end speedup. Its optimization forward pass is $1.87\times$ faster, while its optimization backward pass gives a $4.33\times$ speedup. With Clarabel, the mean time decreases from $38.7$ to $10.8$~min/epoch, corresponding to a $3.58\times$ speedup. Its optimization forward and backward passes are accelerated by $1.99\times$ and $5.17\times$, respectively. In addition, Table~\ref{tab:closed_loop_resource_performance} reports the detailed closed-loop computational performance. Even with the same solver settings as CvxpyLayers (CS mode), the DiffAPQP achieves $1.14\times$ and $1.57\times$ optimization-forward speedups over CvxpyLayers with SCS and Clarabel, respectively as the epigraphical reformulation is avoided (see Section~\ref{sec:disentable_forward_backward}). 

The computational distinction is more pronounced in the counterfactual formulation. CvxpyLayers returns the optimal ED decision variables, from which the ED objective needs to be reconstructed. Computing the counterfactual value loss therefore requires differentiation through the complete ED solution map. DiffAPQP instead can use its value-function layer, which directly returns the optimal ED value. Its gradient with respect to the NN parameter is evaluated from the optimal primal and dual solutions using the envelope-theorem expression from Proposition~\ref{prop:subgradient_qbh} and free-differentiation strategy \eqref{eq:pso_differentiation_free} with negligible backward pass time. This is evidenced by Fig.~\ref{fig:sim_endtoend_counter}. For SCS, the mean recorded time gives a $3.62\times$ speedup. The optimization forward pass is $3.19\times$ faster, while the optimization backward time decreases from $308.6$ to negligible $3.0$~s/epoch. With Clarabel, the mean time produces a \(4.38\times\) speedup. Its optimization backward time decreases from $333.2$ to $6.0$~s/epoch. 

\subsection{Effect of Reduced-KKT Differentiation}

In Table~\ref{tab:closed_loop_resource_performance}, the Clarabel configurations isolate the effect of the reduced-KKT formulation. Changing from the full KKT system to the reduced system reduces the optimization-backward time from \(12.82\) to \(5.47\) minutes, a \(2.34\times\) improvement. In reduced-KKT setting, inactive complementarity equations are removed before solving the adjoint system, which is computationally efficient for the large ED and RD problems considered here.

Moreover, the timing decomposition shows that CvxpyLayers differentiation can be more expensive than solving the optimization problems themselves. CvxpyLayers operates on the conic canonicalization of the parameterized problem and differentiates the corresponding cone-program solution map. In contrast, DiffAPQP instead preserves the APQP structure \eqref{eq:apqp} and computes the required vector--Jacobian product through a sparse adjoint reduced-KKT system \eqref{eq:diff_kkt_reduced}. The results indicate that exploiting this QP structure is particularly beneficial during backpropagation.

\subsection{Effect of Warm-start and Solver Update}

Warm-start and solver-data update affect the optimization-forward pass. DiffAPQP stores solver-native iterates using the unique sample index, allowing each daily sample to reuse its previous solution when it is encountered in a subsequent epoch. However, warm-start and update are not universally beneficial and their effectiveness depends on the algorithm implemented by the solver backend. 

For SCS under the closed-loop setting in Table~\ref{tab:closed_loop_resource_performance}, WS accelerates the optimization forward pass from \(43.54\) to \(26.63\) minutes, corresponding to a \(1.64\times\) speedup. Solver-data update provides no additional benefit: UP is \(1.04\times\) slower than CS in average epoch time, while WS is \(1.24\times\) faster than WU. OSQP benefits more consistently from both mechanisms. Relative to the \(33.57\)-minute CS configuration, UP, WS, and WU achieve \(1.26\times\), \(1.52\times\), and \(1.77\times\) speedups, respectively, with WU producing the fastest OSQP configuration at \(18.96\) minutes per epoch. Since both SCS and OSQP employ operator-splitting algorithms, these results demonstrate the importance of the initial iterate, although the benefit of solver-data update remains backend and problem dependent.

For QPALM, WS reduces the optimization-forward time from \(9.81\) to \(4.16\) minutes and the average epoch time from \(18.08\) to \(10.92\) minutes. Update alone provides little improvement, and combining it with warm-start increases the epoch time to \(16.00\) minutes. Gurobi without warm-start gives the lowest recorded epoch time, \(9.88\) minutes, followed closely by reduced-KKT Clarabel with warm-start at \(10.79\) minutes and QPALM with warm-start at \(10.92\) minutes. For Gurobi, enabling the generic backend warm-start does not help and increases its epoch time to \(10.61\) minutes. Among the open-source backends, Clarabel and QPALM therefore provide performance close to the commercial Gurobi backend.

\subsection{Resource Usage and Decision Quality}\label{sec:simulation_decision_quality}

DiffAPQP also substantially reduces peak memory. 
For the fastest shared-solver configurations, this represents memory reductions of \(55.1\%\) for SCS and \(49.4\%\) for Clarabel, compared to CvxpyLayers where the same solvers are used. The reduction is consistent with retaining the sparse QP representation rather than maintaining the larger conic differentiation representation and its associated derivative state. 

In addition, the test costs obtained using the DiffAPQP differ from those of CvxpyLayers by at most $0.2\%$ using SCS and Clarabel under both closed-loop and counterfactual formulation. The maximum differences along the counterfactual training trajectories are also small, reaching \(0.127\%\) for SCS and \(0.057\%\) for Clarabel. As a result, the computational gains do not materially alter the learning trajectories or test performance. In addition, the close CPU cores usage demonstrate that the observed improvements therefore arise primarily from the solver and differentiation mechanisms rather than from consuming more computational resources.





\section{Conclusion}

This paper presented DiffAPQP, a solver-flexible framework for scalable exact decision-focused learning for large-scale power systems. DiffAPQP automatically canonicalizes optimization models written in CVXPY, accelerates repeated forward solves through solver warm starts and solver-data updates, and supports efficient implicit differentiation through either a full or reduced KKT system which eliminates inactive inequality constraints. For counterfactual value-based learning, we further developed an envelope-theorem-based gradient that completely avoids the time-consuming adjoint KKT solve. Given matched solver configuration, the training performance on IEEE 118-bus system DfL demonstrates that DiffAPQP can achieve \(2.27\times\)--\(3.58\times\) closed-loop and \(3.62\times\)--\(4.38\times\) counterfactual DfL training speedups over CvxpyLayers, while reducing peak memory usage by approximately \(50\%\).


\newpage

\appendix

The appendix provides extra theoretical proofs, experiment setting and results. It starts by proofs on the equivalence between the full-KKT and reduced-KKT system in QP (Proposition~\ref{prop:kkt_equivalence}) and LP (Proposition~\ref{prop:kkt_equivalence_lp}) case, respectively in solution-map layer, as well as the equivalence to the value-function layer differentiation (Proposition~\ref{prop:subgradient_qbh}) in Appendix~\ref{app:proof}; Implementation details of DiffAPQP are explained in Appendix~\ref{app:implementation} with Python pseudo-code for both solution-map and value-function usages;
Extensive experiment settings including reproducibility, power system optimization settings, data preprocessing, neural forecaster structure, training configurations, optimization solver settings, hardware configurations and resource reporting of the DfL training on ED-RD examples in Appendix~\ref{app:exp_setting}; Appendix~\ref{app:counter_result} discusses the extra results on counterfactual DfL training beyond Section~\ref{sec:simulation}; Gradient fidelity between DiffAPQP and CvxpyLayers are summarized in Section~\ref{app:gradient}.

Beyond the experiment on power system ED-RD cases, Appendix~\ref{app:random_exp_setting} initiates experiment on random batched QP and LP studies; Appendix~\ref{app:solver_inference} discuss the detailed setting in customized CVXPY, compared to solver-native interface, original CVXPY, and CvxpyLayers; Appendix~\ref{app:custom_cvxpy} then demonstrates the alignment of DiffAPQP's customized CVXPY with the original CVXPY; Lastly, the performances on random QP/LP studies, including warm-start/solver-update ablation, end-to-end computational time, gradient fidelity, as well as CPU and memory resources are reported in Appendix~\ref{app:random_end_to_end}.

\subsection{Proofs}\label{app:proof}

\subsubsection{Proof to Proposition~\ref{prop:kkt_equivalence}}

First, $K_{a}$ and $D_{a,i}$ in \eqref{eq:diff_kkt_full} can be denoted as
\begin{subequations}
    \begin{equation*}
        K_{a} = 
    \begin{pmatrix}
    P & A^T & G_{\mathcal{A}}^T & G_\mathcal{I}^T  \\
    A & 0 & 0 & 0 \\
    \operatorname{dg}(\tilde{\lambda}_\mathcal{A}^\star)G_\mathcal{A} & 0 & \cancelto{0}{\operatorname{dg}(G_\mathcal{A} \tilde{z}^\star-h_\mathcal{A})} & 0 \\
    \cancelto{0}{\operatorname{dg}(\tilde{\lambda}_\mathcal{I}^\star)G_\mathcal{I}} & 0 & 0 & \operatorname{dg}(G_\mathcal{I} \tilde{z}^\star-h_\mathcal{I})
    \end{pmatrix}
    \end{equation*}
    \begin{equation*}
        \quad D_{a,i} = \begin{pmatrix}
    -Q_i \\ B_i \\ \dg{(\lambda_{\mathcal{A}})H_{i, \mathcal{A}}} \\ \cancelto{0}{\dg(\lambda_\mathcal{I})H_{i,\mathcal{I}}}
\end{pmatrix}
\end{equation*}
\end{subequations}
by setting $\tilde{\lambda}_\mathcal{I}^\star = 0$ and $G_\mathcal{A} \tilde{z}^\star-h_\mathcal{A} = 0$.
Define the vector $\delta = (\delta z^T, \delta \nu^T, \delta \lambda_\mathcal{A}^T, \delta \lambda_\mathcal{I}^T)^T$. We prove that the solution $K_{a}\delta = 0$ is $\delta = 0$. 
Taking the last row-block of $K_{a}$, we have $ \dg(G_\mathcal{I}\tilde{z}^\star - h_\mathcal{I})\delta\lambda_\mathcal{I} = 0$. On the inactive inequality constraint set $\mathcal{I}$, because $(G\tilde{z}^\star - h)_\mathcal{I} < 0$, it gives $\delta \lambda_\mathcal{I} = 0$. For the third row-block, on $\mathcal{A}$, it becomes $G_\mathcal{A}\delta z = 0$ due to assumption (A2). Combining the second row-block in $K_{a}$, e.g., $A\delta z = 0$, it gives that $\delta z\in \operatorname{Null}((A^T, G_\mathcal{A}^T)^T)$. Next, consider the first row-block of $K_{a}$, e.g., $P\delta z + A^T \delta \nu + G_\mathcal{A}^T\delta \lambda_\mathcal{A} + G_\mathcal{I}^T\delta \lambda_\mathcal{I} = 0$. If it is left multiplied by $\delta z^T$, this condition is simplified into
\begin{equation*}
    \delta z^T P\delta z + \cancelto{0}{\delta z^TA^T\delta \nu} + \cancelto{0}{\delta z^T G_\mathcal{A}^T\delta \lambda_\mathcal{A}} + \cancelto{0}{\delta z^T G_\mathcal{I}^T\delta \lambda_\mathcal{I}} = 0
\end{equation*}
Then, because of (A3), $\delta z^T P\delta z=0$ iff $\delta z = 0$. Then the first row-block of $K_{a}$ becomes $A^T \delta \nu + G_\mathcal{A}^T\delta \lambda_\mathcal{A} + \cancelto{0}{G_\mathcal{I}^T\delta \lambda_\mathcal{I}} = 0$. Finally, by (A4), both $\delta \nu$ and $\delta \lambda_\mathcal{A}$ are 0. Consequently, $K_{a}$ is non-singular.

Then consider solving the linear system \eqref{eq:diff_kkt_full}. Similarly as before, the last row-block $\dg{(G_\mathcal{I}\tilde{z}^\star - h_\mathcal{I})}\mathrm{D}\tilde{\lambda}_\mathcal{I}^{a} = 0$ immediately results in $\mathrm{D}\tilde{\lambda}_\mathcal{I}^{a} = 0$ and \eqref{eq:diff_kkt_full} is reduced into
\begin{subequations}\label{eq:diff_kkt_reduce_1}
\begin{equation*}
    K_{2}\cdot\mathrm{D}_{\hat{y}_i}(\tilde{z}^\star, \tilde{\nu}^\star,\tilde{\lambda}_{\mathcal{A}}^\star) = D_{2},\quad \text{with}
\end{equation*}
\begin{equation*}
    K_{2} = \begin{pmatrix}
    P & A^T & G_{\mathcal{A}}^T \\
    A & 0 & 0 \\
    \dg{(\tilde{\lambda}_\mathcal{A}^\star)}G_{\mathcal{A}} & 0 & 0
\end{pmatrix}, ~ D_{2} = \begin{pmatrix}
    -Q_i \\
    B_i \\
    \dg{(\tilde{\lambda}_\mathcal{A}^\star)}H_{i,\mathcal{A}}
\end{pmatrix}
\end{equation*}
\end{subequations}
Because $\lambda_\mathcal{A} > 0$ by (A2), $\dg(\lambda_\mathcal{A})$s are canceled so that $K_{2}$ and $D_{2}$ become $K_{s}$ and $D_{s}$ in \eqref{eq:diff_kkt_reduced}. Then by (A3) and (A4), $K_{s}$ is non-singular so that \eqref{eq:diff_kkt_reduced} has unique solution. Because \eqref{eq:diff_kkt_reduced} is derived from \eqref{eq:diff_kkt_full}, their unique solutions must be the same. Therefore, the full and reduced KKT systems \eqref{eq:diff_kkt_full} and \eqref{eq:diff_kkt_reduced} are equivalent.

\subsubsection{Proof to Proposition~\ref{prop:kkt_equivalence_lp}}

The proof to the LP case is similar to the QP case. First, let $P=0$ in $K_{a}$. Define the vector $\delta = (\delta z^T, \delta \nu^T, \delta \lambda^T_\mathcal{A}, \delta \lambda^T_\mathcal{I} )^T$. We prove that the solution $K_{a}\delta = 0$ is $\delta = 0$. Consider (A5). Being a vertex (basic feasible solution) means that $\operatorname{rank}\left((A^T,G_\mathcal{A}^T)^T\right) = n_d$. The vertex being non-degenerate means that the number of active constraints is $n_d$. Therefore, $(A^T, G_\mathcal{A}^T)^T \in\mathbb{R}^{n_d\times n_d}$ is square and full rank. Then from the last row-block, it immediantely gives that $\delta \lambda_\mathcal{I} = 0$ and the first row-block becomes $A^T\delta \nu + G_\mathcal{A}^T \delta \lambda_\mathcal{A} = 0$. Therefore, $\delta \nu = 0$ and $\delta \lambda_\mathcal{A} = 0$. The second and third row-blocks result in $A\delta z = 0$ and $\dg{(\tilde{\lambda}_\mathcal{A}^\star)}  G_\mathcal{A}\delta z = G_\mathcal{A}\delta z = 0$ due to (A2). As a result, $\delta z = 0$. Consequently, we have proved that $\delta = 0$ for $K_a\delta = 0$.

Then consider solving the linear system \eqref{eq:diff_kkt_full}. It is straightforward to show $\mathrm{D}\tilde{\lambda}_\mathcal{I}^{a} = 0$ and \eqref{eq:diff_kkt_full} is reduced into
\begin{subequations}\label{eq:diff_kkt_reduce_lp}
\begin{equation*}
    K_{3}\cdot\mathrm{D}_{\hat{y}_i}(\tilde{z}^\star, \tilde{\nu}^\star, \tilde{\lambda}_{\mathcal{A}}^\star) = D_{3},\quad \text{with}
\end{equation*}
\begin{equation*}
    K_{3} = \begin{pmatrix}
    0 & A^T & G_{\mathcal{A}}^T \\
    A & 0 & 0 \\
    G_{\mathcal{A}} & 0 & 0
\end{pmatrix}, D_{3} = \begin{pmatrix}
    -Q_i \\
    B_i \\
    H_{i,\mathcal{A}}
\end{pmatrix}
\end{equation*}
\end{subequations}
Because $(A^T, G_\mathcal{A}^T)^T$ is square and full rank, $K_3$ is non-singular and all the conclusion stated in Proposition~\ref{prop:kkt_equivalence_lp} follows.

\subsubsection{Proof to Proposition~\ref{prop:subgradient_qbh}}

To start, the Danskin's theorem \cite{blondel2024elements} is given as follows.

\begin{theorem}
    Let $f: \mathcal{X} \times \mathcal{Y} \rightarrow \mathbb{R}$ and $\mathcal{X}$ be a compact convex set. Let 
    \begin{equation*}
        h(y):=\max_{x \in \mathcal{X}} f(x, y)
    \end{equation*}
    \begin{equation*}
        {x}^{\star}({y}):=\underset{x \in \mathcal{X}}{\arg \max } f(x, y)
    \end{equation*}
If $f$ is concave in $x$ and convex in $y$, and the maximum $x^\star(y)$ is unique, then the function $h$ is differentiable with gradient
\begin{equation*}
    \mathrm{D} h(y)=\mathrm{D}_2 f\left(x^{\star}(y), y\right)
\end{equation*}
If the maximum is not unique, we get a subgradient.
\end{theorem}

Then for Proposition~\ref{prop:subgradient_qbh} with fixed $(b,h)$,
\begin{equation*}
    \alpha(q,c,b,h) = \min_{\tilde z\in\mathcal F(b,h)}
    (\tilde{z}^TP\tilde{z} + q^T\tilde z + c),
\end{equation*}
and
\begin{equation*}
    Z^\star(q,c,b,h) = \arg\min_{\tilde z\in\mathcal F(b,h)}
    (\tilde{z}^TP\tilde{z} + q^T\tilde z + c)
\end{equation*}
where $\mathcal F(b,h):=\{\tilde z\mid A\tilde z=b,\;G\tilde z\le h\}$. 

First, $\alpha(q,c,b,h)=c+\min_{\tilde z\in\mathcal F(b,h)}
(\tilde{z}^TP\tilde{z} + q^T\tilde z)$. Then it is straightforward to see that $\alpha(c)$ is affine on $c$ for some fixed $\tilde{z}$, $b$, and $h$, and
\begin{equation*}
    \mathrm{D}_{c}\alpha(q,c,b,h) = 1
\end{equation*}

For every fixed $\tilde{z}$, $\frac{1}{2}\tilde{z}^TP\tilde{z} + q^T\tilde{z} + c$ is affine in $q$. The point-wise minimum of affine functions $\alpha(\cdot)$ is therefore concave on $q$. Moreover, since $\tilde{z}^TP\tilde{z}+q^T \tilde z + c$ is convex in $\tilde{z}$ and concave in $q$, the Danskin's theorem is applicable such that
\begin{equation*}
    \partial_q^{+}\alpha(q,b,h)=Z^{\star}(q,b,h).
\end{equation*}
Hence, if $Z^{\star}(q,b,h)=\{\tilde z^\star\}$, then $\alpha$ is differentiable with respect to $q$ and
\begin{equation*}
    D_q\alpha(q,b,h)=\tilde z^\star.
\end{equation*}

The convexity on $b$ and $h$ can be directly obtained from perturbation analysis \cite{boyd2004convex}. In addition, the Lagrangian of \eqref{eq:apqp} is
$\mathcal L(\tilde z,\tilde\lambda,\tilde\nu,q,h,b,c)
    =
    \tilde{z}^TP\tilde{z} + q^T \tilde z + c
    +\tilde\lambda^T(G \tilde z-h)
    +\tilde\nu^T(A\tilde z-b)$. Define the dual function
$\phi(\tilde\lambda,\tilde\nu)
    :=
    \min_{\tilde z}
    \{
        \tilde{z}^TP\tilde{z} + q^T\tilde z
        +\tilde\lambda^T G\tilde z
        +\tilde\nu^T A\tilde z
    \} -\tilde{\lambda}^Th - \tilde{\nu}^Tb + c$,
which is concave in $(\tilde{\lambda},\tilde{\nu})$ and convex (and affine) in $(h,b)$. In addition, since Slater's condition holds, strong duality gives
\begin{equation}\label{eq:value_function_dual_form_qbh}
    \alpha(q,b,h)
    =
    \sup_{\tilde\lambda\ge 0,\tilde\nu}
        \phi(\tilde\lambda,\tilde\nu),
\end{equation}
and
\begin{equation*}
    \Lambda^\star(q,b,h) = \sup_{\tilde\lambda\ge 0,\tilde\nu}
        \phi(\tilde\lambda,\tilde\nu).
\end{equation*}
Applying Danskin's theorem to \eqref{eq:value_function_dual_form_qbh} yields
\begin{equation*}
    \partial_{(b,h)}\alpha(q,c,b,h)
    =
    \{(-\nu,-\lambda):(\nu,\lambda)\in\Lambda^{\star}(q,b,h)\}.
\end{equation*}
Hence, if $\Lambda^{\star}(q,b,h)=\{(\tilde\nu^\star,\tilde\lambda^\star)\}$, then $\alpha$ is differentiable with respect to $(b,h)$ and
\begin{equation*}
    D_b\alpha(q,b,h)=-\tilde\nu^\star,
    ~~
    D_h\alpha(q,b,h)=-\tilde\lambda^\star.
\end{equation*}

\subsection{Implementation}\label{app:implementation}





It is worth noting that DiffAPQP's forward solve relies on external solver backends, accessed through CVXPY. As shown in Table~\ref{tab:compare}, although DiffAPQP is general, five solver backends are evaluated in this paper, including operator splitting (OS)-based algorithms OSQP \cite{stellato2020osqp} and SCS \cite{o2021operator}; interior point (IP)-based algorithms Clarabel \cite{gaulart2024clarabel} and Gurobi \cite{gurobi2025}; Augmented Lagrangian (AL)-based algorithms QPALM \cite{hermans2022qpalm}. Meanwhile, the availability of solver-data update and warm start is constrained by both the native solver API and the corresponding CVXPY interface. Some solvers do not expose these features at the solver level; others do, but the current CVXPY interface does not fully expose them.

To address these limitations, we refactor CVXPY's \texttt{solve\_via\_data} routine for OSQP, QPALM, and SCS to decouple the warm-start and update triggers.
Moreover, we do not save the solver caches for each solver, which is memory intensive. Instead, we cache solver workspaces for samples in the first batch of the first epoch, which are then \emph{reused} in all subsequent batches and epochs as discussed in Section~\ref{sec:solver_update}. 
In parallel, we maintain a second solution cache of size $|\mathcal{D}|$, which stores the latest solution for each sample and reuses it as the initial point whenever that sample reappears in following epoch. Consequently, besides the forecast parameters passed to DiffAPQP, a persistent sample index must additionally be provided when sample-indexed explicit warm starting is enabled. For the remaining solvers in CVXPY, we retain the default CVXPY workflow and, when warm-start is triggered, reuse the cached solver instances managed by the original CVXPY.

To sum up, the developed DiffAPQP package implements both solution-map (with full-KKT and reduced-KKT differentiation) and value-function layers for training closed-loop and counterfactual power system DfLs. And the following Python example integrates DiffAPQP into an end-to-end PyTorch training pipeline. Users can specify a power-system optimization problem directly in CVXPY's modeling language and embed it as a differentiable layer. Each layer is initialized once, while the prediction model generates batch-specific optimization parameters. Persistent dataset indices associate each sample with its cached solver state, enabling subsequent calls to reuse previous solutions through warm starting; problem-data updates avoid repeated solver initialization. In the meantime, DiffAPQP supports parallel forward solves and backward differentiation, which is essential for stochastic training with large sample batches. 

The solution-map layer returns decision variables as a dictionary indexed by their names defined when constructing the corresponding CVXPY optimization model. The value function layer directly outputs the optimal objective. In both cases, gradients propagate through the optimization problem by a counterfactual MSE loss \eqref{eq:counterfactual_dfl} to update the prediction model using standard PyTorch operations. As the input parameters are also named, the gradient with respect to the output of the forecasters can be easily obtained.

\begin{python}
from diffapqp import ValueFuncLayer, SolMapLayer
# Initialize: prob is cvxpy.Problem instance.
layer_val = ValueFuncLayer(prob) 
layer_sol = SolMapLayer(prob)
# Sample index for warm-start from cached sol.
idx_list = sample_idx.detach().cpu().tolist()
# Forward pass via torch.nn.Module predictor.
load_hat, solar_hat = nn(x)
param_dict = {"load_hat": load_hat,
              "solar": solar_hat}
solver = "OSQP"
# Value-function layer.
c_hat, *_ = layer_val(
    param_dict, idx_list=idx_list,
    solver=solver, solver_args=solver_args,
    warm_start=True, update=True
)
loss=0.5/x.shape[0]*torch.square(c_hat - c).sum()
# Alternatively, use the solution-map layer.
primal_dict, *_ = layer_sol(param_dict,
    idx_list=idx_list, solver=solver,
    solver_args=solver_args, warm_start=True,
    update=True, reduced_kkt=True
)
dispatch_hat = primal_dict["dispatch"]
c_hat = cost_fn(dispatch_hat)
loss=0.5/x.shape[0]*torch.square(c_hat - c).sum()
# Backward pass and parameter update.
optimizer.zero_grad()
loss.backward()
\end{python}

\subsection{Extensive Experiment Settings of Section~\ref{sec:simulation} }\label{app:exp_setting}

The detailed experiment settings for Section~\ref{sec:simulation} is described in this appendix.

\subsubsection{Reproducibility}

The DiffAPQP and benchmark on random batched QP/LP experiment in Appendix~\ref{app:random_exp_setting}--\ref{app:random_end_to_end} are available at \url{https://github.com/xuwkk/diffapqp} and can be installed by \texttt{pip install diffapqp}. The ED-RD experiment in Section~\ref{sec:simulation} is reproducible at \url{https://github.com/xuwkk/diffapqp_power_systems}. The modified version of CVXPY is maintained separately at \url{https://github.com/xuwkk/cvxpy} based on CVXPY v1.9.0.

\subsubsection{Power System Decision Making Settings}

The performance of the proposed DiffAPQP framework is compared with CvxpyLayers on the IEEE 118-bus system under closed-loop and counterfactual DfL settings, as shown in Fig.~\ref{fig:dfl}. The test system contains 118 buses, 34 generators, 186 transmission branches, 99 loads, 10 solar plants, and 10 wind plants. The ED and RD models are continuous convex quadratic programs incorporating generator capacity and ramping constraints, DC nodal power balance, transmission limits, renewable curtailment, load shedding, and emergency generation shedding. The RD model additionally represents asymmetric upward and downward redispatch costs.

A 24-hour optimization horizon is used for both ED and RD. The ED and RD problems contain 8,952 and 10,584 scalar decision variables, respectively. Both models contain 4,488 scalar equality constraints, while ED and RD contain 19,536 and 24,432 scalar inequality constraints, respectively. This problem size provides a large-scale stress test for differentiable optimization in end-to-end DfL, which is rare in the current research.

\subsubsection{Data Acquisition and Preprocessing}

One year of hourly nodal load, solar, wind, and weather data is obtained from the quasi-realistic Texas system~\cite{lu2025synthetic}. The IEEE 118-bus testbed, profile assignment, and data preprocessing are implemented using our open-source GridForge package~\cite{xu2025lapso}. The forecaster inputs comprise cyclical weekday and hour encodings, temperature, shortwave and longwave radiation, zonal and meridional wind speeds, and total wind speed. Non-calendar features and forecast targets are standardized, while the predicted load, solar, and wind profiles are transformed back to MW and constrained to be nonnegative before being passed to the optimization layers.

The dataset contains 8,737 rolling 24-hour windows and 365 non-overlapping daily windows. The daily windows are randomly divided into 255 training, 54 validation, and 56 test samples. Rolling windows are used for AbL, whereas the non-overlapping daily windows are used for DfL fine-tuning.

\subsubsection{Neural Forecaster Structure}

We employ a shared multi-task neural forecasting model to jointly predict the 24-hour trajectories of system loads, solar generation, and wind generation. The model consists of three components: a normalized spatiotemporal input representation, a shared two-dimensional convolutional encoder, and three asset-specific forecasting heads. By sharing the encoder across forecasting tasks, the model can exploit common temporal and exogenous patterns while retaining separate output mappings for different asset types.

Let the normalized input context window be denoted by
\begin{equation*}
    {X}
    \in
    \mathbb{R}^{B \times T \times N_{bus} \times N_{feature}}
\end{equation*}
where $B$ is the batch size, $T=24$ is the number of hourly time steps, $N_{bus}=118$ is the number of buses, and $N_{feature}=10$ is the number of input features. The feature set contains four calendar variables, one temperature variable, two solar-radiation variables, and three wind-speed variables. The input is rearranged into the channel-first format required by the convolutional layers:
\begin{equation*}
    \widetilde{{X}}
    =
    \operatorname{permute}({X})
    \in
    \mathbb{R}^{B \times N_{feature} \times T \times N_{bus}}
\end{equation*}

The shared encoder contains two consecutive two-dimensional convolutional layers with ReLU activations:
\begin{align*}
     {H}^{(1)}
    &=
    \sigma\!\left(
    \operatorname{Conv2D}_{10 \rightarrow 128}
    (\widetilde{{X}})
    \right) \in
    \mathbb{R}^{B \times 128 \times 24 \times 118} \\
    {H}^{(2)}
    &=
    \sigma\!\left(
    \operatorname{Conv2D}_{128 \rightarrow 256}
    ({H}^{(1)})
    \right) \in
    \mathbb{R}^{B \times 256 \times 24 \times 118}
\end{align*}
where $\sigma(\cdot)$ denotes the ReLU activation function. Both convolutional layers use a kernel size of $3 \times 5$ and padding of $1 \times 2$ which preserves the temporal and bus dimensions. The convolutional kernels jointly process the time and bus axes, allowing the encoder to capture short-term temporal dependencies and correlations between nearby buses under the adopted bus ordering. 

To obtain a compact system-level representation, the encoded features are averaged along the bus axis:
\begin{equation*}
    \textstyle {H}_{b,c,t}
    =
    \frac{1}{N_{bus}}
    \sum_{n=1}^{N_{bus}}
    {H}^{(2)}_{b,c,t,n} \in
    \mathbb{R}^{B \times 256 \times 24}
\end{equation*}
Hence, each forecasting hour is represented by a 256-dimensional latent vector that aggregates information from the entire network. 

Three independent prediction heads are then applied to the shared latent representation. Each head consists of a one-dimensional convolution with kernel size one. The forecasting heads are defined as
\begin{align*}
    \widetilde{{Y}}_{{load}}
    &=
    \operatorname{Conv1D}_{256 \rightarrow 99}
    ({H}) \\
    \widetilde{{Y}}_{{solar}}
    &=
    \operatorname{Conv1D}_{256 \rightarrow 10}
    ({H}) \\
    \widetilde{{Y}}_{{wind}}
    &=
    \operatorname{Conv1D}_{256 \rightarrow 10}
    ({H})
\end{align*}

After transposing the channel and time dimensions, the final normalized forecasts are
\begin{equation*}
    \widehat{{Y}}_{{load}}\in\mathbb{R}^{B \times 24 \times 99}, \widehat{{Y}}_{{solar}}\in\mathbb{R}^{B \times 24 \times 10}, \widehat{{Y}}_{{wind}}\in
    \mathbb{R}^{B \times 24 \times 10}
\end{equation*}
Finally, the normalized outputs are transformed back to physical units in MW using the corresponding inverse normalization. Since negative load and renewable-generation values are physically infeasible, the reported forecasts and the values passed to downstream optimization are projected onto the nonnegative domain:
\begin{equation*}
    \widehat{{Y}} =
    \max\left\{
    \operatorname{Denorm}
    (\widehat{{Y}}),
    0
    \right\}
\end{equation*}
Overall, the architecture combines parameter-efficient shared representation learning with task-specific output layers, thereby capturing common weather and temporal drivers while preserving the distinct forecasting characteristics of load, solar, and wind assets.

\subsubsection{Training Configurations}

The AbL forecaster is trained with Adam by minimizing the equally weighted mean of the normalized load, solar, and wind MSEs. We perform a grid search over$\texttt{hidden\_dim}\in\{128,256\}, \texttt{num\_layers}\in\{2,3\},\quad
\texttt{lr}\in\{0.001,0.005,0.01\}, \texttt{batch\_size}\in\{16,32,64\}$, giving 36 configurations. Each configuration is trained for at most 100 epochs, with early stopping after 10 validation epochs without improvement. The checkpoint with the lowest validation normalized MSE is selected. The resulting forecaster uses $\texttt{hidden\_dim}=256$, two convolutional layers, a learning rate of $0.005$, and a batch size of 32. The selected model contains 541,687 trainable parameters.

The selected AbL checkpoint initializes all DfL experiments. Fine-tuning uses Adam with a learning rate of $10^{-5}$ and a daily batch size of 32. Five epochs are recorded, where the first epoch evaluates the initial ABL model without parameter updates and the subsequent four epochs perform DfL fine-tuning. In the closed-loop setting, the predicted profiles are first passed through ED. The resulting generation schedule is then passed to RD together with the realized load, solar, and wind profiles. Training minimizes the mean realized end-to-end RD cost, scaled by $10^{6}$ for numerical stability during backpropagation, while the reported operational cost remains in dollars. In the counterfactual setting, training minimizes the loss \eqref{eq:counterfactual_dfl}. The true ED costs are computed using Gurobi and cached for all daily samples.

In DiffAPQP, warm starts are indexed using each sample's unique start-hour identifier, allowing a daily problem to retrieve its previous solution across epochs.

\subsubsection{Solver Arguments}

The solver arguments used for Section~\ref{sec:simulation} are reported in Table~\ref{tab:solver_argument}. Default arguments are used if not specified in the table in other simulations in Appendix.

\begin{table*}[t]
    \centering
    \caption{Solver and their algorithm types considered in the paper. Default settings are used if not specified.}
    \footnotesize
    \begin{tabular}{cccc}
    \textbf{Solver}   &  \textbf{Prob. Type} & \textbf{Algorithm} & \textbf{Settings} \\
    \midrule
    \textbf{OSQP} & QP & Operator splitting (ADMM) & \texttt{eps\_abs}:1e-4, \texttt{eps\_rel}:1e-4, \texttt{max\_iter}: 10,000 \\
    \textbf{SCS} & CP & Operator splitting & \texttt{eps\_abs}:1e-5, \texttt{eps\_rel}:1e-5, \texttt{max\_iters}:100,000 \\
    \textbf{QPALM} & QP & Augmented-Lagrangian & \texttt{eps\_abs}:1e-6, \texttt{eps\_rel}:1e-6, \texttt{max\_iter}: 10,000 \\
    \textbf{Gurobi} & QP & Gurobi chooses automatically & \texttt{Threads}:1, \texttt{OptimalityTol}:1e-8, \texttt{FeasibilityTol}:1e-8, \texttt{BarConvTol}: 1e-8 \\
    \textbf{Clarabel} & CP & Interior point & \texttt{tol\_gap\_abs}:1e-8, \texttt{tol\_gap\_rel}:1e-8, \texttt{max\_iter}:10,000 \\
    \bottomrule
    \end{tabular}
    \label{tab:solver_argument}
\end{table*}

The primary closed-loop DiffAPQP experiments use reduced-KKT differentiation with MINRES and a relative tolerance of $10^{-9}$. Full-KKT experiments use LSQR with absolute and relative tolerances of $10^{-6}$ and a maximum of 5,000 iterations. For each solver backend, DiffAPQP and CvxpyLayers use identical solver tolerances and training configurations.

\subsubsection{Hardware Configuration}\label{app:cpu_restriction}

All DfL timing experiments are executed on a dual-socket workstation equipped with two Intel Xeon Gold 6230R CPUs at 2.10\,GHz and 256\,GB of system memory whereas AbL pretraining is performed using an NVIDIA GeForce RTX 3090 GPU. All methods use the same random seed, data split, forecaster initialization, batch size, learning rate, and solver arguments whenever the same backend is used.

To represent a stress test, the number of CPU cores are controlled as 32 for both CvxpyLayers and DiffAPQP training with the following prefix for all the experiment: \texttt{taskset -c 0-31}.

\subsubsection{CPU and Memory Usage Reporting}\label{app:usage_reporting}

We record CPU and memory usage with a lightweight sampler running at 0.05 s intervals. CPU usage is measured for all recursive child processes, so multiprocessing workers are included correctly. The CPU usage is reported as the mean effective number of cores, computed from process-level CPU utilization during the measured interval (including both forward and backward passes).
Memory usage is measured from system-wide used memory via \texttt{psutil.virtual\_memory().used}. The memory sampler starts after benchmark data generation and before layer construction, so memory monitoring is end-to-end, including layer construction, solver/cache initialization, the base solve, perturbed solve, and backward pass. Memory usage is reported as the peak increase in system memory during the same interval.

Wall-clock time is decomposed into neural-network forward time, optimization forward time, optimization-layer backward time, and neural-network backward time. First-call setup and canonicalization overhead are retained in the reported optimization-forward time. Resource usage is sampled every 50\,ms for the training process and its child processes. CPU utilization is reported as the average effective number of occupied cores, computed from process CPU time relative to wall-clock time. Memory usage is reported as the peak increase in system-wide memory relative to the baseline measured at the beginning of each batch. All reported experiments use the fixed random seed 404.

\subsection{Simulation Results for Counterfactual DfL in Power System}\label{app:counter_result}

\begin{table*}[t]
    \centering
    \caption{Counterfactual DfL performance for all evaluated solver configurations. The overall epoch time includes the recorded training
    forward/backward and validation forward components, whereas the optimization
    component columns include training only.}
    \label{tab:counterfactual_resource_performance}
    \small
    \begin{adjustbox}{width=\textwidth}
    \begin{tabular}{
        l l c c
        S[table-format=2.2]
        S[table-format=2.2]
        S[table-format=2.2]
        S[table-format=1.3]
        S[table-format=2.2]
        S[table-format=1.8]
    }
    \toprule
    \textbf{Method}
    & \textbf{Solver}
    & \textbf{W}
    & \textbf{U}
    & {\textbf{Avg. cores}}
    & {\textbf{Peak memory (GiB)}}
    & {\textbf{Train opt. fwd. (min)}}
    & {\textbf{Train opt. bwd. (min)}}
    & {\textbf{Avg. epoch time (min)}}
    & {\textbf{Test MSE $(\$/10^6)^2$}} \\
    \midrule
    
    CvxpyLayers & SCS & -- & -- &
    15.87 & 38.68 & 27.16 & 5.144 & 38.16 & 0.00985687 \\
    DiffAPQP & SCS & False & False &
    19.82 & 11.93 & 23.23 & 0.111 & 28.16 & 0.00985561 \\
    DiffAPQP & SCS & False & True &
    20.82 & 11.96 & 22.06 & 0.047 & 26.46 & 0.00985482 \\
    DiffAPQP & SCS & True & False &
    10.58 & 11.94 & 8.51 & 0.049 & 10.53 & 0.00985908 \\
    DiffAPQP & SCS & True & True &
    9.04 & 11.94 & 9.43 & 0.112 & 11.80 & 0.00986004 \\
    
    \addlinespace
    CvxpyLayers & Clarabel & -- & -- &
    17.26 & 34.46 & 4.11 & 5.553 & 10.73 & 0.00985635 \\
    DiffAPQP & Clarabel & False & {N/A} &
    22.60 & 12.86 & 2.33 & 0.052 & 3.13 & 0.00985667 \\
    DiffAPQP & Clarabel & True & {N/A} &
    24.47 & 12.89 & 1.73 & 0.099 & 2.45 & 0.00985667 \\
    
    \addlinespace
    DiffAPQP & OSQP & False & False &
    21.24 & 12.53 & 7.93 & 0.054 & 9.82 & 0.00985253 \\
    DiffAPQP & OSQP & False & True &
    19.90 & 12.51 & 8.44 & 0.055 & 9.97 & 0.00984188 \\
    DiffAPQP & OSQP & True & False &
    11.69 & 12.50 & 1.84 & 0.100 & 2.63 & 0.00986297 \\
    DiffAPQP & OSQP & True & True &
    13.46 & 12.49 & 2.09 & 0.048 & 2.80 & 0.00986361 \\
    
    \addlinespace
    DiffAPQP & QPALM & False & False &
    22.45 & 13.45 & 3.63 & 0.050 & 4.71 & 0.00985682 \\
    DiffAPQP & QPALM & False & True &
    21.36 & 13.44 & 3.69 & 0.055 & 4.82 & 0.00985682 \\
    DiffAPQP & QPALM & True & False &
    13.21 & 13.45 & 1.11 & 0.098 & 1.68 & 0.00985671 \\
    DiffAPQP & QPALM & True & True &
    17.35 & 13.50 & 2.25 & 0.057 & 3.01 & 0.00985685 \\
    
    \addlinespace
    DiffAPQP & Gurobi & False & {N/A} &
    9.36 & 12.41 & 1.51 & 0.090 & 2.11 & 0.00985667 \\
    DiffAPQP & Gurobi & True & {N/A} &
    7.98 & 12.39 & 1.81 & 0.047 & 2.46 & 0.00985667 \\
    
    \bottomrule
    \end{tabular}
    \end{adjustbox}
\end{table*}

As a complementary to Table~\ref{tab:closed_loop_resource_performance}, Table~\ref{tab:counterfactual_resource_performance} shows a larger DiffAPQP advantage in the counterfactual formulation. With SCS, warm-started DiffAPQP reduces the mean epoch time from \(38.16\) to \(10.53\) minutes, giving a \(3.62\times\) speedup. Its optimization-forward pass is \(3.19\times\) faster, while optimization-backward time decreases from \(5.144\) to \(0.049\) minutes, corresponding to an approximately \(104\times\) speedup. With Clarabel, DiffAPQP reduces the epoch time from \(10.73\) to \(2.45\) minutes, giving a \(4.38\times\) speedup. Its forward and backward components are accelerated by \(2.38\times\) and \(56\times\), respectively.

The particularly large backward speedup follows from the DiffAPQP value-function mechanism. CvxpyLayers reconstructs the optimal ED cost from the primal solution and differentiates through the complete conic solution map. DiffAPQP instead evaluates the value gradient directly from the optimal primal and dual variables using the envelope theorem, avoiding an adjoint KKT solve. Consequently, the DiffAPQP backward time is only a few seconds per epoch and is negligible relative to its forward optimization time.

Warm-start provides the most consistent forward-pass improvement. It accelerates SCS, OSQP, and QPALM by \(2.67\times\), \(3.73\times\), and \(2.80\times\), respectively, relative to their cold-start configurations. Solver-data update provides little additional benefit and generally makes the warm-started configurations slower. The fastest DiffAPQP configuration is warm-started QPALM at \(1.68\) minutes per epoch. Compared with the fastest CvxpyLayers configuration, Clarabel at \(10.73\) minutes, this represents a cross-backend speedup of \(6.39\times\).

DiffAPQP also reduces peak memory from \(38.68\) to \(11.94\)~GiB with SCS and from \(34.46\) to \(12.89\)~GiB with Clarabel, corresponding to reductions of \(69.1\%\) and \(62.6\%\). These gains do not materially affect decision quality: the SCS and Clarabel test-MSE differences relative to CvxpyLayers are only \(0.022\%\) and \(0.003\%\), respectively.

\subsection{Gradient Agreement and Training-Path Divergence in Power Systems}\label{app:gradient}

\begin{table*}[t]
    \centering
    \caption{Per-sample cosine similarity between DiffAPQP and CvxpyLayers forecast gradients for the \emph{first} batch of the \emph{first} DfL training epoch. Both methods use Clarabel, and all 32 samples are evaluated before the first optimizer update. Solar and wind statistics include only positive forecasts.}
    \label{tab:gradient_similarity_first_batch}
    \small
    \begin{adjustbox}{width=\textwidth}
    \begin{tabular}{
        l l l
        S[table-format=2.0]
        S[table-format=1.6] S[table-format=1.6]
        S[table-format=1.6] S[table-format=1.6]
        S[table-format=1.6] S[table-format=1.6]
    }
    \toprule
    \textbf{DiffAPQP layer}
    & \textbf{Formulation}
    & \textbf{Solver}
    & {\textbf{Samples}}
    & \multicolumn{2}{c}{\textbf{Load}}
    & \multicolumn{2}{c}{\textbf{Solar \(>0\)}}
    & \multicolumn{2}{c}{\textbf{Wind \(>0\)}} \\
    \cmidrule(lr){5-6}
    \cmidrule(lr){7-8}
    \cmidrule(lr){9-10}
    &&&&
    {\textbf{Mean}} & {\textbf{P05}}
    & {\textbf{Mean}} & {\textbf{P05}}
    & {\textbf{Mean}} & {\textbf{P05}} \\
    \midrule
    Full KKT
    & Closed-loop
    & Clarabel
    & 32
    & 0.983099 & 0.942964
    & 0.980163 & 0.940726
    & 0.981288 & 0.942180 \\
    
    Reduced KKT
    & Closed-loop
    & Clarabel
    & 32
    & 0.999541 & 0.997859
    & 0.999324 & 0.997350
    & 0.999484 & 0.997752 \\
    
    Value function
    & Counterfactual
    & Clarabel
    & 32
    & 0.999897 & 0.999705
    & 0.999916 & 0.999849
    & 0.999978 & 0.999921 \\
    \bottomrule
    \end{tabular}
    \end{adjustbox}
\end{table*}

\begin{table*}[t]
    \centering
    \caption{Cosine similarity over \emph{all} training samples stored during the \emph{first} DFL training epoch. Both methods use Clarabel. Unlike Table~\ref{tab:gradient_similarity_first_batch}, later batches are evaluated after method-specific optimizer updates and gradient mismatches are expected.}
    \label{tab:gradient_similarity_all_samples}
    \small
    \begin{adjustbox}{width=\textwidth}
    \begin{tabular}{
        l l l
        S[table-format=3.0]
        S[table-format=1.6] S[table-format=1.6]
        S[table-format=1.6] S[table-format=1.6]
        S[table-format=1.6] S[table-format=1.6]
    }
    \toprule
    \textbf{DiffAPQP layer}
    & \textbf{Formulation}
    & \textbf{Solver}
    & {\textbf{Samples}}
    & \multicolumn{2}{c}{\textbf{Load}}
    & \multicolumn{2}{c}{\textbf{Solar \(>0\)}}
    & \multicolumn{2}{c}{\textbf{Wind \(>0\)}} \\
    \cmidrule(lr){5-6}
    \cmidrule(lr){7-8}
    \cmidrule(lr){9-10}
    &&&&
    {\textbf{Mean}} & {\textbf{P05}}
    & {\textbf{Mean}} & {\textbf{P05}}
    & {\textbf{Mean}} & {\textbf{P05}} \\
    \midrule
    Full KKT
    & Closed-loop
    & Clarabel
    & 255
    & 0.974053 & 0.929482
    & 0.960763 & 0.872009
    & 0.974022 & 0.928346 \\
    
    Reduced KKT
    & Closed-loop
    & Clarabel
    & 255
    & 0.991010 & 0.986669
    & 0.980531 & 0.898744
    & 0.989846 & 0.984164 \\
    
    Value function
    & Counterfactual
    & Clarabel
    & 255
    & 0.998134 & 0.995841
    & 0.713828 & 0.050653
    & 0.965518 & 0.692364 \\
    \bottomrule
    \end{tabular}
    \end{adjustbox}
\end{table*}

In Section~\ref{sec:simulation_decision_quality}, it has been demonstrated that various forward and backward settings result in similar training and test losses compared to CvxpyLayers' training outcome. To verify this observation, Tables~\ref{tab:gradient_similarity_first_batch} and
\ref{tab:gradient_similarity_all_samples} compare the gradients of DiffAPQP against those of CvxpyLayers with respect to the load, solar, and wind forecasts, under the closed-loop and counterfactual DfL settings in Section~\ref{sec:simulation}. The closed-loop comparison evaluates the full- and reduced-KKT solution-map layers, while the counterfactual comparison evaluates the value-function layer. Clarabel is used for both methods. The cosine similarity is first computed for each daily sample by flattening its 24-hour forecast gradient, after which the mean and fifth percentile (P05) are reported.

During DfL training, solar and wind forecasts are restricted to positive values before passing to the ED layers. At a zero renewable forecast, the lower and upper curtailment bounds are simultaneously active, e.g., $0 \leq p^{\mathrm{curt}}_{\mathrm{solar}} \leq \widehat{p}_{\mathrm{solar}}=0$, 
resulting in a degenerate KKT point at which the derivative need not be
unique. CvxpyLayers and DiffAPQP may consequently select different numerical
derivatives at these boundary points. Moreover, gradients associated with
forecasts clipped from negative values are blocked by the clamp and do not
propagate to the NN. The positive-forecast comparison therefore better
represents the effective differentiable training signal.

Table~\ref{tab:gradient_similarity_first_batch} provides the controlled
layer-to-layer comparison. All 32 samples in this batch are evaluated using
the same pretrained AbL forecaster before either method performs an optimizer
update. The full-KKT layer achieves mean cosine similarities of \(0.9831\),
\(0.9802\), and \(0.9813\) for load, solar, and wind, respectively. Its P05
values remain above \(0.9407\), showing that the gradients are strongly
aligned for most samples. The residual discrepancy is consistent with the
approximate solution of the large full asymmetric adjoint system \eqref{eq:diff_kkt_full} using LSQR with finite termination tolerances.

The reduced-KKT formulation \eqref{eq:diff_kkt_reduced} improves the agreement substantially. Its mean
cosine similarities are \(0.99954\), \(0.99932\), and \(0.99948\), while all
P05 values exceed \(0.9973\). Thus, even the lower tail of the sample-wise
distribution remains closely aligned with CvxpyLayers. Under the evaluated
configuration, this improvement reflects the combined effect of removing
inactive constraints from the adjoint system to obtain more robust results and solving the resulting smaller system using MINRES with a tighter tolerance. 

The value-function layer provides the closest agreement, with mean cosine
similarities of \(0.99990\), \(0.99992\), and \(0.99998\) for load, solar, and
wind, respectively. Its P05 values are also above \(0.9997\). The nearly
identical gradients demonstrate that, at regular positive-forecast points,
the proposed value-function formulation (Proposition~\ref{prop:subgradient_qbh} and \eqref{eq:pso_differentiation_free}) can reproduce the CvxpyLayers training signal
without requiring an adjoint KKT solve.

Table~\ref{tab:gradient_similarity_all_samples} aggregates the 255 common
training samples stored after the first DfL training epoch. The full-KKT
mean similarities remain between \(0.9608\) and \(0.9741\), while the
reduced-KKT values remain between \(0.9805\) and \(0.9910\). However, this
table is not a controlled measure of intrinsic layer fidelity. After the
first batch, CvxpyLayers and DiffAPQP update their forecasters independently.
Consequently, the gradients for subsequent samples are evaluated at different
NN parameters, forecast values, and optimization active sets, even though there disagreements are expected and the training losses are close.

This distinction is particularly important for the value-function solar result. Its all-sample mean decreases to \(0.7138\), compared with \(0.99992\) in the controlled first batch before training. Solar forecasts frequently lie near the clipping boundary because of their diurnal profile. Small method-specific parameter updates can therefore move a forecast across zero, changing the active constraints discontinuously. In addition, the positive-renewable mask used for the all-sample analysis is reconstructed from the initial AbL forecaster and may no longer match the evolving forecasts in later batches. Accordingly, Table~\ref{tab:gradient_similarity_first_batch} is used as the primary gradient-fidelity result, whereas Table~\ref{tab:gradient_similarity_all_samples} is interpreted only as a diagnostic of divergence along the sequential training path.

\subsection{Experiment Settings on Random Quadratic and Linear Programs in the Appendix}
\label{app:random_exp_setting}

\begin{table*}[t]
    \centering
    \footnotesize
    \caption{Optimization data generation used in the random QP /LP benchmarks.}
    \begin{tabularx}{\linewidth}{Xp{2.8cm}p{5.8cm}p{3.5cm}p{3.8cm}}
    \toprule
    & \multicolumn{2}{c}{\textbf{Dense}} & \multicolumn{2}{c}{\textbf{Sparse}} \\
    & \textbf{Random Matrix} & \textbf{Well-conditioned Matrix} & \textbf{Random Matrix} & \textbf{Well-conditioned Matrix} \\
    \midrule
    $P$
    & 1. Sample $P_r$ with iid standard normal entries. \newline
      2. Set $P=P_r^\top P_r + \epsilon I$.
    & 1. Sample $P_r$ with iid standard normal entries. \newline
      2. Compute QR factorization: $P_r=\mathcal{Q}_P\mathcal{R}_P$. \newline
      3. Set eigenvalues $\lambda_i$ evenly in $[1,10]$. \newline
      4. Set $P=\mathcal{Q}_P \dg(\lambda)\mathcal{Q}_P^\top$, so that $\kappa(P)=10$.
    & 1. Sample $z_i \sim \mathcal{N}(0,1)$. \newline
      2. Set $P=\dg(0.5+|z_i|)$.
    & 1. Set diagonal entries $\lambda_i$ evenly in $[1,10]$. \newline
      2. Set $P=\dg(\lambda)$, so that $\kappa(P)=10$.
    \\
    \midrule
    $A$
    & Sample $A$ with iid standard normal entries.
    & 1. Sample $A_r$ with iid standard normal entries. \newline
      2. Compute QR factorization: $A_r^\top=\mathcal{Q}_A\mathcal{R}_A$. \newline
      3. Set $A=\mathcal{Q}_A^\top$, so that the rows of $A$ are orthonormal and $\kappa(A)=1$.
    & Apply an elementwise Bernoulli mask with density $0.1$ to the dense random $A$.
    & Apply an elementwise Bernoulli mask with density $0.1$ to the dense well-conditioned $A$.
    \\
    \midrule
    $G$
    & Sample $G$ with iid standard normal entries.
    & 1. Sample $G_r$ with iid standard normal entries. \newline
      2. Normalize each row of $G_r$ to unit norm to obtain $G$.
    & Apply an elementwise Bernoulli mask with density $0.1$ to the dense random $G$.
    & Apply an elementwise Bernoulli mask with density $0.1$ to the dense row-normalized $G$.
    \\
    \bottomrule
    \end{tabularx}
    \label{tab:data_generation}
\end{table*}

This section summarizes the data generation and solver arguments for \emph{random QP and LP experiment in the Appendix}. Note that the settings in this section \emph{may not be followed in the ED-RD examples} in Section~\ref{sec:simulation}.

\subsubsection{Data Generation}

We benchmark end-to-end performance of DiffAPQP (both solution map Layer and value function layer) with CvxpyLayers for both QP and LP. The problem size is denoted as $n_d$, $n_{in}$, and $n_{eq}$, respectively. 

The following arguments are considered:
\begin{itemize}
    \item solver = \{OSQP, SCS, QPALM, Clarabel, Gurobi\} for DiffAPQP and solver = \{SCS, Clarabel\} for CvxpyLayers.
    \item matrix\_format = \{dense, sparse\}.
    \item problem\_type = \{QP, LP\}.
    \item data\_mode = \{random, well\_conditioned\}. 
    \item For DiffAPQP, consider \{cold-start (CS), with warm-start only (WS), with update only (UP), with both warm-start and update (WU)\}.
    \item For DiffAPQP with Layer setting, consider \{with and without reduced adjoint system formulation\}. The LSQR and MINRES solvers are used for full and reduced adjoint system, respectively.
    \item For DiffAPQP with Layer setting under LP, we test regularizer within \{1e-3,0\}. 
\end{itemize}

The generation procedures for optimization data $P,A,G$ under different problem settings are summarized in Table~\ref{tab:data_generation}, which are used for all samples in a batch. Moreover, the condition number of $Q$, $A$, and $G$ matrices not only influence the efficiency of the forward solution, but also impact on the solution quality of the adjoint system and the stability of the reduced-KKT setting in the backward differentiation. 

\subsubsection{Parameter Generation}

For each fixed problem template $(P,A,G)$, we generate \emph{batches} of vector parameters while keeping the matrix data fixed. For each sample, we generate a \emph{base} instance $(q,h,b)$ and a \emph{perturbed} instance $(q',h',b')$. It is assumed that $P$, $A$, and $G$ have been generated based on Table~\ref{tab:data_generation}.

\paragraph{QP Problem}

For the QP setting, the main design focus is to be \emph{feasible}. To start, the linear cost vectors are sampled independently as $q_i \sim \mathcal{N}(0,I), ~ i=1,\dots,N$ where $N$ represents the batch size. Then, for each sample, we draw a reference primal point $x_i^{\mathrm{ref}} \sim \mathcal{N}(0,I)$. The equality right-hand side is chosen so that $x_i^{\mathrm{ref}}$ is feasible: $b_i = A x_i^{\mathrm{ref}}$. The inequality right-hand side is generated as $h_i = G x_i^{\mathrm{ref}} + s_i$, where the slack vector $s_i$ is strictly positive. In the implementation, the slack is sampled elementwise as $s_i = 0.5 + 0.2 |z_i|, ~ z_i \sim \mathcal{N}(0,1)$, so that $x_i^{\mathrm{ref}}$ is \emph{strictly feasible} for the inequality constraints.

To build the perturbed dataset, we generate a nearby point $x_i^{\mathrm{pert}} = x_i^{\mathrm{ref}} + \sigma \xi_i,
~ \xi_i \sim \mathcal{N}(0,I)$, where $\sigma > 0$ is the perturbation scale. The perturbed equality and inequality right-hand side is $b_i' = A x_i^{\mathrm{pert}}$ and $h_i' = G x_i^{\mathrm{pert}} + s_i$ where $s_i$ is the same as the base case. The cost vector is kept unchanged: $q_i' = q_i$.

\paragraph{LP Problem}

For the LP setting, the main design purpose is to have \emph{bounded} solution. To start, the cost vector is not sampled independently. Instead, it is constructed so that a sampled primal point satisfies the KKT conditions. First, we draw a reference primal point $x_i^{\mathrm{ref}} \sim \mathcal{N}(0,I)$, and define the equality right-hand side by $b_i = A x_i^{\mathrm{ref}}$. Next, an active-set mask is sampled for the inequality constraints. A slack vector $s_i$ is generated such that active constraints have zero slack and inactive constraints have strictly positive slack. The inequality right-hand side is then $h_i = G x_i^{\mathrm{ref}} + s_i$, with

\begin{equation*}
(s_i)_j =
\begin{cases}
0, & j \in \mathcal{A}_i, \\
>0, & j \in \mathcal{I}_i.
\end{cases}
\end{equation*}
A nonnegative dual vector $\lambda_i$ is sampled so that $\lambda_i \ge 0, (\lambda_i)_j = 0 ~ \text{for } j \in \mathcal{I}_i$, and an equality multiplier $\nu_i$ is sampled from a Gaussian distribution. The cost vector is then defined through the stationarity condition:
\begin{equation*}
q_i = -(A^\top \nu_i + G^\top \lambda_i)
\end{equation*}
With this construction, the tuple $(x_i^{\mathrm{ref}}, \lambda_i, \nu_i)$ satisfies primal feasibility, dual feasibility, complementarity, and stationarity. Therefore, $x_i^{\mathrm{ref}}$ is an optimal solution of the generated LP instance.

The perturbed LP data is generated following the same procedure as QP.


\paragraph{Differentiable Loss}

The optimal objective value $L^\star = \frac{1}{2}x^{\star,T} P x^\star + q^Tx^\star$ is used as the scalar loss and differentiated with respect to the problem parameters. DiffAPQP's value function layer directly returns $L^\star$, enabling the ``differentiation-for-free'' strategy. In contrast, the DiffAPQP solution-map layer and CvxpyLayers return the optimal decision variable $x^\star$; consequently, the objective must first be reconstructed from $x^\star$ before differentiation.

\subsection{Comparison of Solver-Interface Pipelines}\label{app:solver_inference}

\begin{table*}[t]
    \centering
    \footnotesize
    \caption{Warm-start and update behaviors in solver-native interface, default CVXPY interface, custom CVXPY interface and CvxpyLayers interface on selected APQP solution algorithms. The custom CVXPY behaviors only support APQP solution.}
    \begin{tabularx}{\linewidth}{Xp{5.6cm}p{4.5cm}p{3cm}p{1.8cm}}
    \toprule 
    \textbf{Solver} & \textbf{Solver-native Interface} & \textbf{Original CVXPY} & \textbf{Custom CVXPY} & \textbf{CvxpyLayers} \\
    \midrule
    \textbf{OSQP} \cite{stellato2020osqp} & \texttt{solver.warm\_start(x=x0, y=y0)} \newline \texttt{solver.update(q=q\_new, l=l\_new, u=u\_new)} & Both are supported with one \texttt{warm\_start} flag to trigger both. & Separate \texttt{warm\_start} and \texttt{update} flags as in original solver backend. & N/A \\
    \midrule
    \textbf{SCS} \cite{o2021operator} & \texttt{solver.solve(warm\_start=True, x=x0, y=y0, s=s0)} \newline \texttt{solver.update(b=b\_new, c=c\_new)} & Only warm-start is supported by the \texttt{warm\_start} flag & Separate \texttt{warm\_start} and \texttt{update} flags as in original solver backend. & Does not support warm-start and update. \\
    \midrule
    \textbf{QPALM} \cite{hermans2022qpalm} & \texttt{solver.warm\_start(x0, y0)} \newline \texttt{solver.update\_q(q\_new)}, \texttt{solver.update\_bounds(bmin\_new, bmax\_new)} & Both are supported with one \texttt{warm\_start} flag to trigger both. & Separate \texttt{warm\_start} and \texttt{update} flags as in original solver backend. & N/A \\
    \midrule 
    \textbf{Clarabel} \cite{gaulart2024clarabel} 
    & No public API for user-specified initial point. \newline \texttt{solver.update(P=P\_new, q=q\_new, A=A\_new, b=b\_new)}  
    & No initial-point warm-start. The \texttt{warm\_start} flag reuses the cached solver and triggers data update when allowed. & Same as original CVXPY. & Does not support warm-start and update. \\
    \midrule
    \textbf{Gurobi} \cite{gurobi2025} 
    & Continuous QP warm-start can use advanced-start information such as \texttt{PStart}/\texttt{DStart}. \newline
    Existing models can be update in place, e.g., \texttt{setAttr}, \texttt{chgCoeff}, \texttt{setMObjective}, then \texttt{optimize()}.
    & Partial warm-start only: CVXPY rebuilds the Gurobi model and sets variable \texttt{Start} from the cached solution. In-place update is not supported.  & Same as original CVXPY. & N/A \\
    \bottomrule
    \end{tabularx}
    \label{tab:solver_behaviour}
\end{table*}

\begin{figure}[h]
    \centering
    \includegraphics[width=0.85\linewidth]{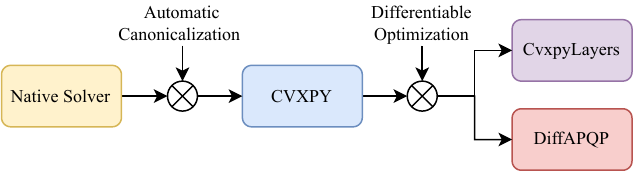}
    \caption{Brief relationships among the compilers considered in the paper. Detailed comparison between CvxpyLayers and DiffAPQP is in Table~\ref{tab:compare}.}
    \label{fig:compiler_compare}
\end{figure}

In this paper, each solver backend can be accessed through four types of solver-interface pipelines: (i) the solver-native interface, (ii) the default CVXPY interface, (iii) the custom CVXPY interface used by DiffAPQP, and (iv) the CvxpyLayers interface. Their relationships are highlighted in Fig.~\ref{fig:compiler_compare}.

The solver-native interface usually requires the optimization problem to be manually converted into the solver's standard form before calling the backend. These standard forms differ across solvers. For example,
\begin{equation}\label{eq:solver_standard_form}
    \begin{aligned}
        \text{OSQP:}~~\min~&\textstyle\frac{1}{2} x^TPx + q^Tx \\
        \text{s.t.} ~ & l\leq Ax\leq u \\
        \text{SCS:}~~\min~&\textstyle\frac{1}{2} x^TPx + c^Tx \\
        \text{s.t.} ~ & Ax+s = b, ~~s\in\mathcal{K} \\
        \text{QPALM:}~~\min~&\textstyle\frac{1}{2} x^TQx + q^Tx \\
        \text{s.t.} ~ & b_{min} \leq Ax \leq b_{max} \\
        \text{Clarabel:}~~\min~&\textstyle\frac{1}{2} x^TPx + q^Tx \\
        \text{s.t.} ~ & Ax + s = b,~~ s\in\mathcal{K}
    \end{aligned}
\end{equation}
Here, we retain each solver's native parameter notation. As discussed in Section~\ref{sec:canonicalization}, manually converting a problem into solver standard form requires solver-specific optimization knowledge and is prone to implementation errors. In contrast, CVXPY, CvxpyLayers, and DiffAPQP interfaces provide automatic canonicalization from a high-level problem description under different settings.

Moreover, as discussed in Section~\ref{sec:solution}, native solver APIs usually expose warm-start and problem-data update mechanisms to different extents. However, these features are not be fully exposed through the default CVXPY interface or CvxpyLayers. The differences are summarized in Table~\ref{tab:solver_behaviour}. For example, SCS natively supports matrix factorization reuse via \texttt{solver.update} and manual warm-start overrides for primal and dual variables, whereas the current CVXPY's SCS interface only passes cached previous solve under \texttt{warm\_start=True} and does not expose the persistent-workspace update path. More generally, in the current CVXPY interfaces for OSQP and QPALM, solver update and warm start are coupled under a single \texttt{warm\_start} flag. CVXPY also documents that warm start uses variable value fields only on the first solve; on subsequent solves, initialization is taken from the cached previous solution \cite{diamond2016cvxpy}; therefore, user cannot pass arbitrary warm start point to the solver. Consequently, a refactor is needed to better control the warm-start and update behaviors in the custom CVXPY-based interface and DiffAPQP. 

At last, as both Clarabel and Gurobi have limited warm-start and update exposure, DiffAPQP uses the default CVXPY's \texttt{warm\_start} flag and does not separate the update function as the other solvers.

\begin{table*}[t]
    \centering
    \footnotesize
    \caption{Performance of default and custom CVXPY interfaces for solving random QP problem. The reported computational time is end-to-end, averaged over solving 5 samples in serial. Numbers in the brackets are the iterations. For each solver, results in same color are comparable as per Table~\ref{tab:solver_behaviour}. }
    \begin{tabularx}{\linewidth}{Xp{2.2cm}p{2.2cm}|p{2.2cm}p{2.2cm}p{2.2cm}p{2.2cm}}
    \toprule
    & \multicolumn{2}{c|}{\textbf{Default CVXPY}} & \multicolumn{3}{c}{\textbf{Custom CVXPY}} \\
    & \textbf{No Warm-start} & \textbf{With Warm-start} & \textbf{Without Both} & \textbf{With Warm-start} & \textbf{With Update} & \textbf{With Both} \\
    \midrule
    \textbf{OSQP} & \textcolor{red}{5.16 (830)} & \textcolor{blue}{2.88 (725)} & \textcolor{red}{4.97 (830)} & 4.36 (725) & 3.34 (830) & \textcolor{blue}{2.84 (725)} \\
    \textbf{SCS} & \textcolor{red}{17.15 (2910)} & \textcolor{blue}{16.17 (2940)} & \textcolor{red}{16.66 (2910)} & \textcolor{blue}{16.29 (2940)} & 12.64 (2640) & 11.83 (2565)  \\
    \textbf{QPALM} & \textcolor{red}{18.22 (115.6)} & \textcolor{blue}{9.94 (161.4)} & \textcolor{red}{18.17 (115.6)} & 9.93 (161.4) & 18.04 (115.6) & \textcolor{blue}{9.94 (161.4)} \\
    \textbf{Clarabel} & \textcolor{red}{3.71 (12.60)} & \textcolor{blue}{3.26 (12.60)} & \textcolor{red}{3.65 (12.60)} & \textcolor{blue}{3.23 (12.60)} & \textcolor{blue}{3.39 (12.60)} & \textcolor{blue}{3.28 (12.60)}  \\
    \textbf{Gurobi} & \textcolor{red}{14.11 (13.60)} & \textcolor{blue}{13.75 (13.60)} & \textcolor{red}{14.01 (13.60)} & \textcolor{blue}{13.73 (13.60)} & \textcolor{blue}{13.76 (13.60)} & \textcolor{blue}{13.73 (13.60)} \\ 
    \bottomrule
    \end{tabularx}
    \label{tab:custom_cvxpy}
\end{table*}

\subsection{Simulation on Default Custom CVXPY Interface}\label{app:custom_cvxpy}

This section demonstrates warm-start and update behavior for OSQP, SCS, QPALM, Clarabel and Gurobi for both default and custom CVXPY interfaces.

\subsubsection{Settings}

The solver arguments are reported in Table~\ref{tab:solver_argument} with \texttt{eps\_abs}=\texttt{eps\_rel}=1e-6 and 1e-5 modified for SCS and OSQP, respectively. The QP case with dense and random matrix setting in Table~\ref{tab:data_generation} is considered with $n_d = 700$, $n_{in} = 1000$, and $n_{eq} = 400$ and the perturbation scale $\sigma=0.1$. Whenever a problem is resolved with warm-start and/or update option, we always reset and solve the baseline problem before passing to the perturbed problem for fair comparison.

\subsubsection{Results}

The results are summarized in Table~\ref{tab:custom_cvxpy}. The computational performances highlighted in the same color represent the same underlying warm-start and update settings as Table~\ref{tab:solver_behaviour}. To start, for all five solvers, original CVXPY without warm-start has close performance of custom CVXPY with no warm-start and update triggered. The same iteration numbers and similar computational performance verify that the custom CVXPY follows the same default CVXPY warm-start behavior where \texttt{update} has no effect in Clarabel and Gurobi solvers as the default CVXPY warm-start is implemented.

Moreover, for SCS solver, the CVXPY \texttt{warm\_start} flag only triggers the warm-start functionality but ignores the update such as matrix factorization caching. This is evidenced by 16.17 s and 16.29 s warm-start solution time in the default and custom CVXPY solution time. In contrast, CVXPY \texttt{warm\_start} for OSQP and QPALM solvers implement both warm-start and update, as confirmed by the similar end-to-end computational time and iterations for custom CVXPY with both \texttt{warm\_start} and \texttt{update} triggers.

In addition, different solvers have different warm-start and update performances on a given optimization problem. For the specific QP settings defined in Appendix~\ref{app:random_exp_setting}, both update and warm-start benefits on the OSQP, and SCS while the improvement from update is more significant. For QPALM, warm-start dominates the computational efficiency improvement while update has little improvement. Note that the reported warm-start and update preference may be different to different QP or LP settings.

\subsection{Performance of DiffAPQP on Batch-solve Random QP and LP}\label{app:random_end_to_end}

This section reports random QP and LP runs with 32 samples in parallel. All settings follow Appendix~\ref{app:custom_cvxpy}. The CPU cores are limited by Appendix~\ref{app:cpu_restriction}.

\subsubsection{Analysis on Warm-start and Update in the Forward Solution}

\begin{figure*}[t]
    \centering
    \includegraphics[width=0.9\linewidth]{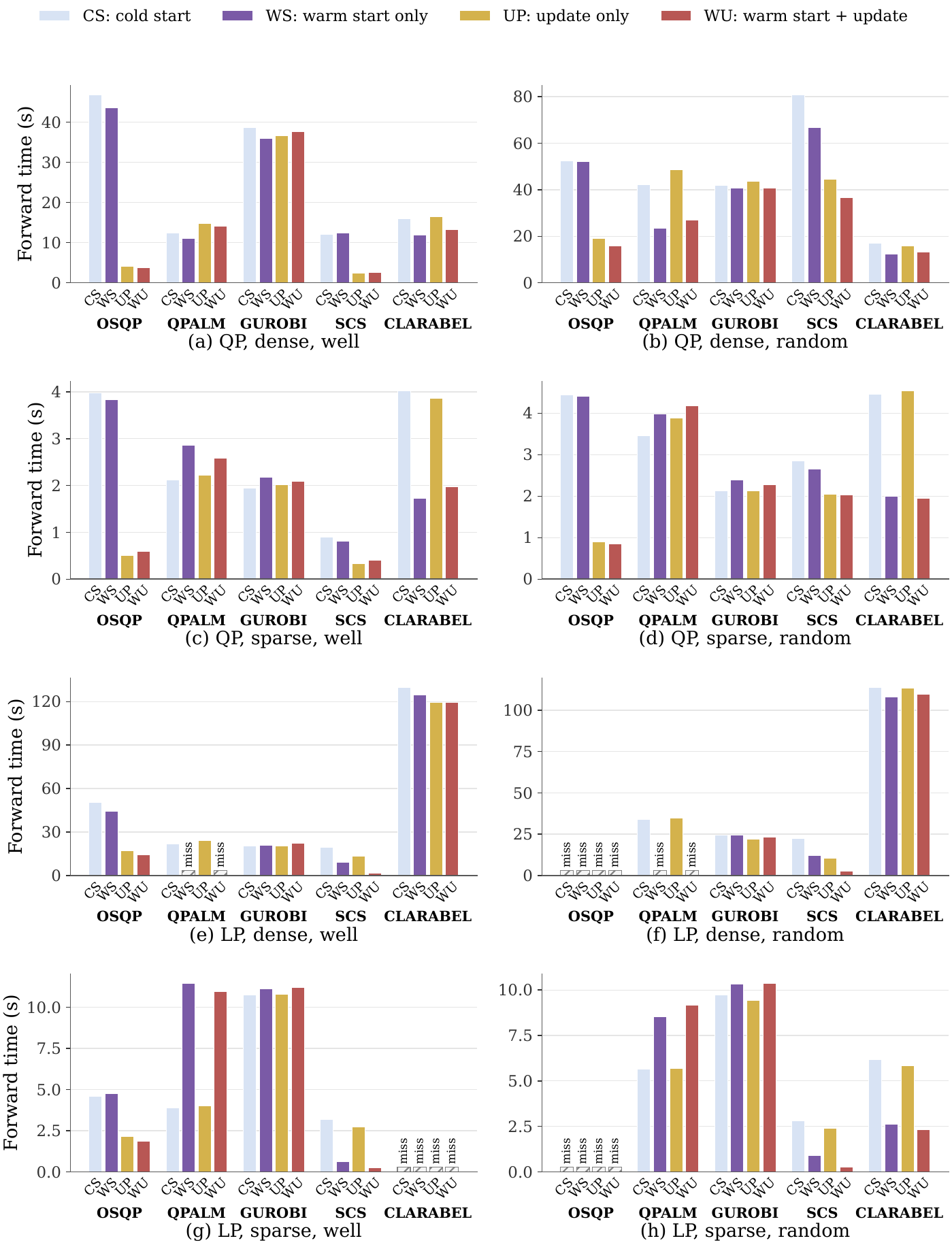}
    \caption{DiffAPQP forward-pass solution time under cold start (CS), warm start only (WS), update only (UP), and combined warm start and update (WU), evaluated across solver backends, QP and unregularized LP problems, dense and sparse matrices, and well-conditioned and random data.}
    \label{fig:forward_mode_compare}
\end{figure*}

We evaluate how warm starting and solver-state updating affect the forward-pass performance of DiffAPQP under four execution modes: cold start (CS), warm start only (WS), update only (UP), and combined warm start and update (WU). The reported time covers the forward evaluation of the \emph{perturbed} batch. Layer initialization and the initial base solve are excluded. Thus, CS denotes a second solve without reusing the previous solution or solver workspace, rather than the first invocation of the layer under base data setting. All modes solve the same generated instances and use the same solver settings in Table~\ref{tab:solver_argument} (with the same modifications in Appendix~\ref{app:custom_cvxpy}), so comparisons between modes are paired within each solver. The performance is reported in Fig.~\ref{fig:forward_mode_compare} where no regularization terms are added on LP problems. 
\begin{tcolorbox}[
    colback=white,
    colframe=black,
    boxrule=0.5pt,
    left=4pt,
    right=4pt,
    top=4pt,
    bottom=4pt
]
The main findings on warm-start and update about different solver backends are listed below:
\begin{enumerate}
    \item No warm-start and update strategy is universally optimal;
    \item OSQP benefits primarily from updating;
    \item SCS benefits most from combining both mechanisms and SCS WU shows particularly strong LP improvements;
    \item Clarabel benefits mainly from solver update;
    \item QPALM's warm-start and update performances vary on different problems;
    \item Gurobi shows little systematic improvement.
\end{enumerate}
\end{tcolorbox}

The explicit DiffAPQP update interface is available for OSQP, QPALM, and SCS in this comparison. Gurobi instead uses its conventional primal warm-start mechanism, while Clarabel's CVXPY warm-start path also reuses and updates its solver workspace. Consequently, the separation between warm start and update should primarily be interpreted for OSQP, QPALM, and SCS.

Solver-state updating is the principal source of acceleration for OSQP. Across the six configurations that OSQP solves successfully, UP and WU achieve geometric-mean speedups of $4.43\times$ and $4.80\times$, respectively, relative to CS, whereas WS alone provides only a $1.04\times$ speedup. Update-enabled modes reduce forward time by 53-92\%, with particularly large improvements for QPs. In detail, the dense well-conditioned QP decreases from 46.82 s under CS to 3.81 s under WU, while the sparse random QP decreases from 4.43 s to 0.85 s. These results indicate that reusing the factorization and solver workspace is considerably more valuable for OSQP than supplying only the previous primal-dual iterate.

SCS exhibits a different and complementary pattern. Although WS and UP independently provide geometric-mean speedups of $1.77\times$ and $1.87\times$, their combination achieves a $5.12\times$ speedup averaged over eight problem settings. The interaction is especially strong for LPs. The WU reduces forward time by 88-92\% in every LP configuration. For example, the dense well-conditioned LP decreases from 19.54 s to 1.54 s, and the sparse random LP decreases from 2.80 s to 0.26 s. The combined improvement is substantially larger than either mechanism alone, suggesting that SCS benefits simultaneously from retaining its workspace and initializing the updated problem with a nearby solution.

The benefits are less universal for the remaining solvers. Clarabel benefits mainly from its warm-start and solver update path, with an aggregate speedup of approximately $1.58\times$. The improvement is strongest for sparse problems, reaching 50-62\%, but is small for dense LPs. QPALM shows no consistent acceleration. Updating alone is close to CS, while warm-start modes can become substantially slower and fail for the dense unregularized LPs. Gurobi also shows no systematic timing improvement because each model is reconstructed and warm starting only supplies an initial primal solution. Its observed differences are generally within the variability expected from independent timing runs.

Overall, the results demonstrate that forward-pass reuse in DiffAPQP must be selected according to the underlying solver, given the specific problem. The 16 missing results, all arising from unregularized LP configurations, also indicate that solver robustness is more sensitive in the LP setting and motivate the $\epsilon=10^{-3}$ regularized term. Nevertheless, the multi-fold OSQP and SCS improvements provide strong evidence for the value of solver-aware warm-start and solver update in DiffAPQP.

\subsubsection{End-to-end Forward and Backward Time Comparison for Solution Map and Value Function Layers}\label{app:end_to_end_compare}

\begin{figure*}[t]
    \centering
    \includegraphics[width=0.9\linewidth]{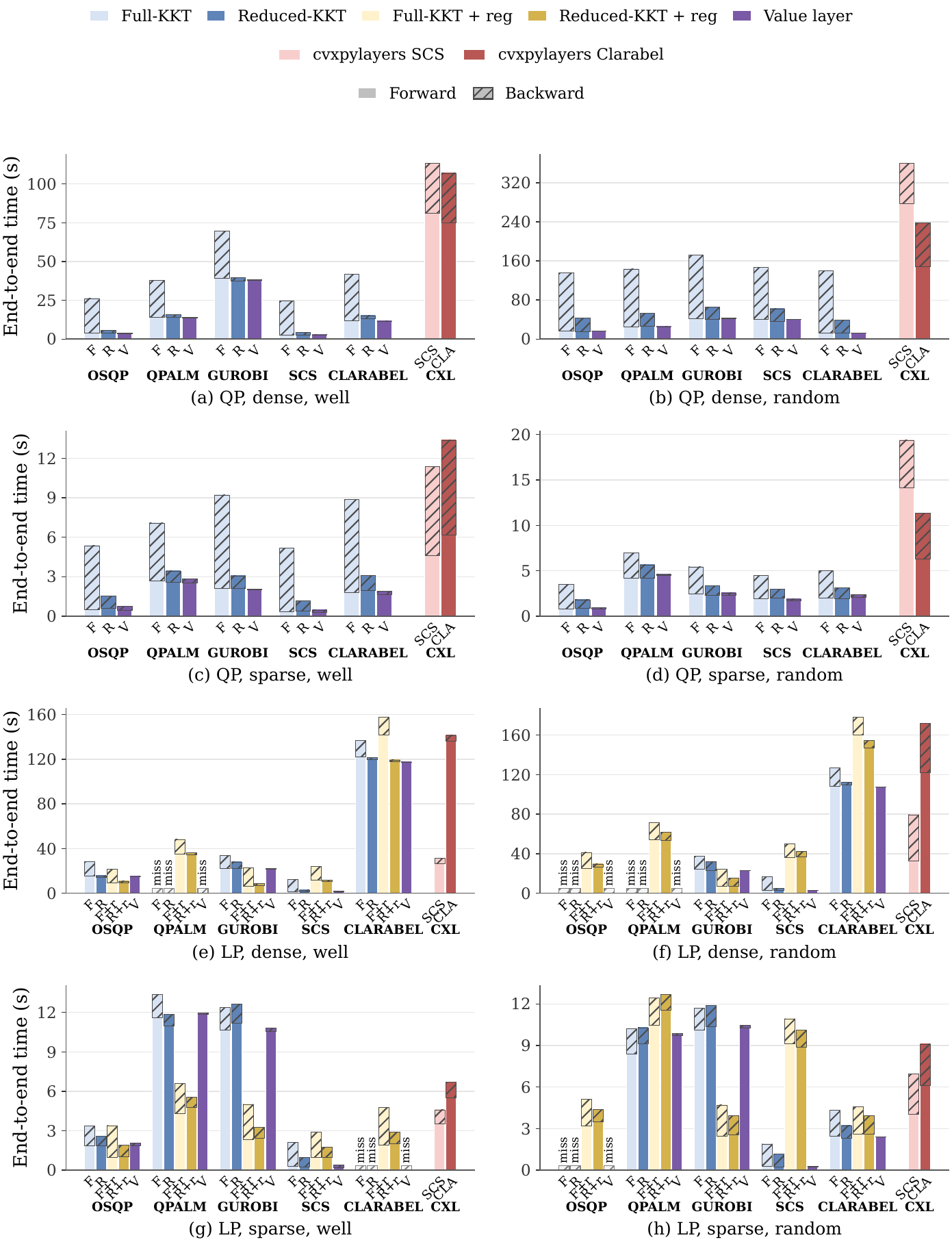}
    \caption{End-to-end runtime of the DiffAPQP solution-map and value-function layers and CvxpyLayers, with stacked bars separating forward and backward time. DiffAPQP uses combined warm start and problem-data update (WU), with solution-map gradients computed using either Full-KKT or Reduced-KKT adjoints. LP results additionally include quadratic regularization ($\epsilon=10^{-3}$). CvxpyLayers results using SCS and Clarabel are shown separately in the CXL group.}
    \label{fig:random_end_to_end_time}
\end{figure*}

We compare the end-to-end computational time of the DiffAPQP solution-map and value-function layers against CvxpyLayers in Fig.~\ref{fig:random_end_to_end_time}. The reported time is the sum of the forward and backward passes for the perturbed batch. Layer construction and the initial base solve are excluded, so the results represent the steady-state performance of repeatedly solving and differentiating related optimization problems. The DiffAPQP time is reported with both warm-start and solver update triggered while CvxpyLayers does not expose equivalent warm-start or update reuse and directly evaluates the perturbed batch. 
\begin{tcolorbox}[
    colback=white,
    colframe=black,
    boxrule=0.5pt,
    left=4pt,
    right=4pt,
    top=4pt,
    bottom=4pt
]
The main findings are summarized as follows:
\begin{enumerate}
    \item Reduced KKT setting consistently improves differentiation efficiency by $4.13\times$ on average, resulting in a $2.02\times$ geometric-mean end-to-end speedup over full KKT differentiation;
    \item Compared to CvxpyLayers, DiffAPQP accelerates the forward pass by $13.16\times$ and $3.00\times$ for SCS and Clarabel solvers in average; 
    \item The value layer is fastest when only objective sensitivities are required.
\end{enumerate}
\end{tcolorbox}

To start, with the same underlying solver, DiffAPQP achieves a geometric-mean forward speedup of $13.16\times$ over CvxpyLayers with SCS across all eight problem settings. The corresponding speedups are $11.84\times$ for QPs and $14.63\times$ for LPs, with individual improvements ranging from $6.99\times$ to $32.46\times$. With Clarabel, DiffAPQP achieves a $3.00\times$ geometric-mean forward speedup across the seven successful settings. The improvement is stronger for QPs, at $5.05\times$, than for LPs, at $1.50\times$. These results show that DiffAPQP's advantage is already substantial before differentiation. It arises from reusing the canonicalized APQP representation, solver workspaces, and previous primal-dual solutions when evaluating a nearby parameter batch.

Full-KKT differentiation constructs the complete complementarity system and solves it using LSQR \cite{paige1982lsqr}, whereas reduced KKT retains only the active inequality constraints and solves the resulting symmetric system using more efficient MINRES \cite{paige1975solution}. The active set is selected using an absolute tolerance of $10^{-3}$ and zero relative tolerance. Among the 35 solver-problem combinations for which both unregularized full-KKT and reduced-KKT runs completed successfully, reduced KKT achieved a geometric-mean end-to-end speedup of $2.02\times$. The backward pass itself is $4.13\times$ faster, and its median share of total runtime decreases from 62\% under full KKT to 32\% under reduced KKT. The largest observed end-to-end improvement is $5.62\times$. Reduced KKT is therefore the preferred DiffAPQP solution-map configuration when sensitivities of the optimizer are required.

Combining the faster forward and backward passes produces substantial solver-matched improvements over CvxpyLayers. With SCS, full-KKT DiffAPQP is $3.17\times$ faster end to end, while reduced-KKT DiffAPQP is $9.07\times$ faster. The reduced-KKT advantage is $9.81\times$ for QPs and $8.40\times$ for LPs, with individual speedups between $4.79\times$ and $25.66\times$. With Clarabel, the corresponding geometric-mean speedups are $1.71\times$ for full KKT and $3.19\times$ for reduced KKT. Reduced-KKT DiffAPQP is particularly effective for Clarabel-based QPs, where it achieves a $5.09\times$ speedup, while the improvement for LPs is $1.71\times$. Because these comparisons use the same solver on the same problem instances, they provide stronger evidence for the benefits of DiffAPQP's repeated-solve and differentiation architecture than comparisons across different solver families.

When only the gradient of the optimal value is required, DiffAPQP's value function layer provides a further reduction in differentiation cost. Instead of differentiating the optimizer through a KKT adjoint system, it applies the envelope theorem directly to the stored primal-dual solution. Across 35 matched configurations, the value layer is $1.50\times$ faster end to end than the reduced-KKT solution-map layer, while its backward pass is $13.85\times$ faster. The median contribution of backward differentiation falls to approximately 2.2\% of total runtime. Relative to solver-matched CvxpyLayers, the value layer achieves geometric-mean end-to-end speedups of $18.20\times$ with SCS and $4.60\times$ with Clarabel. It is the fastest method in all eight problem categories. 

For LPs, we additionally evaluate quadratic regularization with $\epsilon=10^{-3}$. All regularized full- and reduced-KKT runs complete successfully, whereas the figure contains 15 missing unregularized results caused by solver failures. Regularization is especially beneficial for Gurobi, reducing reduced-KKT end-to-end time by $2.06\times$ to $3.85\times$, and it enables several previously failing OSQP, QPALM, and Clarabel cases. However, its performance effect is not universally positive. For example, regularized SCS runs are approximately $1.8\times$ to $8.3\times$ slower than their unregularized counterparts. The reason is explained in Section~\ref{sec:disentable_forward_backward}. Adding second-order regularization term transforms the LP into QP, which requires extra variables and rotated cones in SCS.

\subsubsection{Gradient Agreement and Differentiation Reliability}

\begin{figure*}[t]
    \centering
    \includegraphics[width=0.85\linewidth]{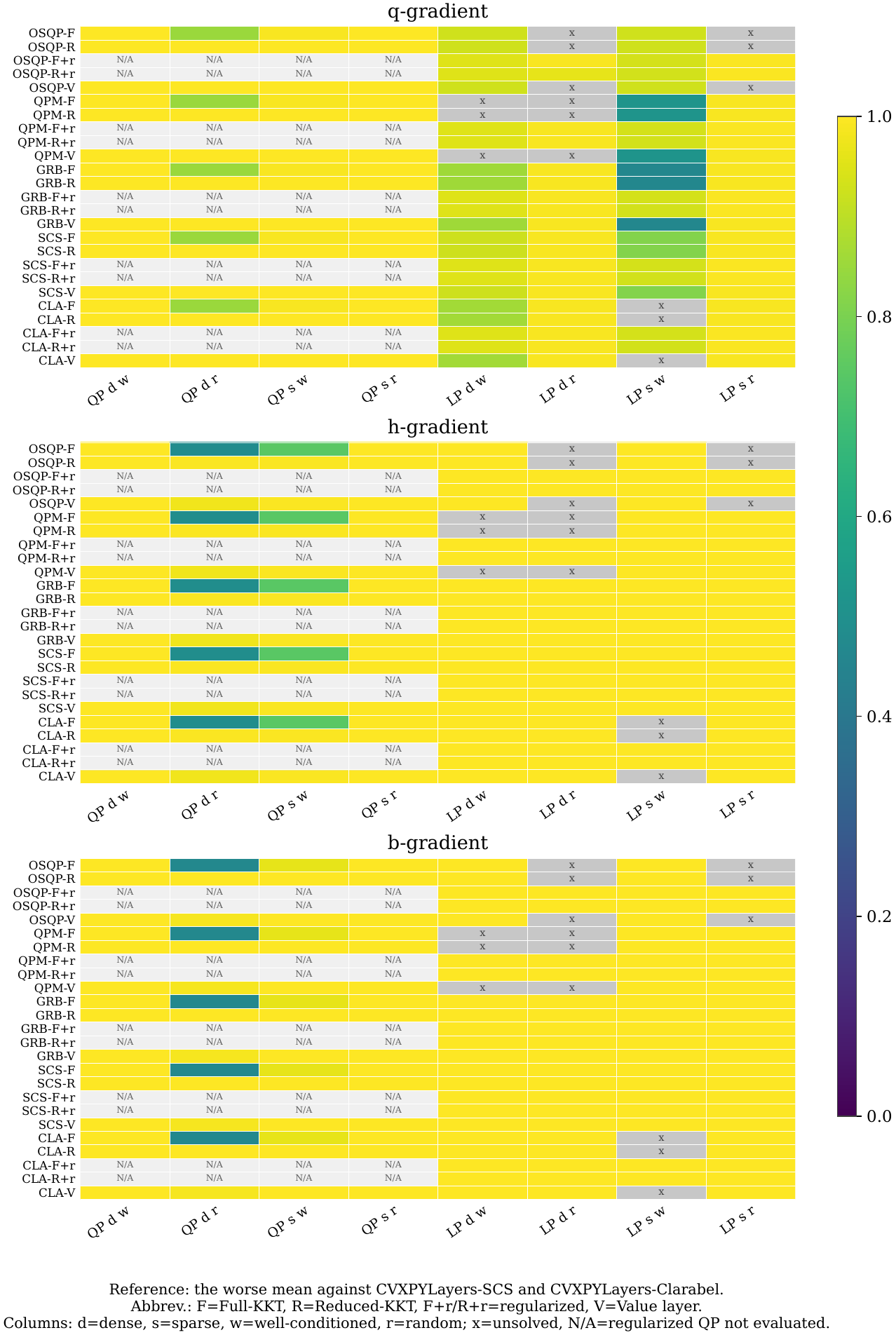}
    \caption{Mean sample-wise cosine similarity between DiffAPQP gradients and CvxpyLayers references. For each gradient component $q$, $h$, and $b$, similarities are averaged over the batch separately against CvxpyLayers-SCS and CvxpyLayers-Clarabel, and the lower of the two means is reported.}
    \label{fig:random_cos_sim}
\end{figure*}

For all the experiments in Appendix~\ref{app:end_to_end_compare}, considering the DiffAPQP gradient component $z\in\{q,h,b\}$ and CvxpyLayers reference solver $r\in\{\mathrm{SCS},\mathrm{Clarabel}\}$, we compute the mean sample-wise cosine similarity
\begin{equation*}
    \textstyle C_z^{(r)} = \frac{1}{B}\sum_{i=1}^{B}
    \frac{\left\langle \nabla_z \widehat{L_i},
    \nabla_z L_i^{(r)}\right\rangle}{
    \|\nabla_z \widehat{L_i}\|_2
    \|\nabla_z L_i^{(r)}\|_2}
\end{equation*}
where $\widehat{(\cdot)}$ represents the loss computed by the DiffAPQP layers. The performance is illustrated in Fig.~\ref{fig:random_cos_sim} with the conservative value $C_z=\min_r C_z^{(r)}$. 

\begin{tcolorbox}[
    colback=white,
    colframe=black,
    boxrule=0.5pt,
    left=4pt,
    right=4pt,
    top=4pt,
    bottom=4pt
]
The following observations are made:
\begin{enumerate}
    \item Reduced-KKT is the most reliable solution-map configuration for QPs; The value function layer achieves similarly strong agreement and is preferable when only optimal-value sensitivities are needed;
    \item Full-KKT differentiation is sensitive to challenging QPs, particularly dense-random problems;
    \item LP $h$- and $b$-gradients are consistently reliable while unregularized LP $q$-gradients are solver dependent;
    \item Quadratic regularization improves LP consistency and solver robustness.
\end{enumerate}
\end{tcolorbox}

The QP results show a particularly clear advantage for reduced-KKT. Across all five solvers and four QP settings, every reduced-KKT cosine similarity exceeds $0.9939$. Its mean/minimum similarities are $1.000/0.9999$ for $q$, $0.997/0.9939$ for $h$, and $0.999/0.9964$ for $b$. The value function layer is similarly reliable, with minimum similarities of $0.9999$, $0.9799$, and $0.9861$, respectively.

In contrast, full-KKT differentiation is less reliable for the more difficult QP settings. For the dense-random QP, its solver-averaged similarities fall to approximately $0.851$, $0.489$, and $0.466$ for the $q$-, $h$-, and $b$-gradients in the worst case. The sparse well-conditioned QP also reduces the full-KKT $h$-gradient similarity to approximately $0.743$. These patterns are nearly identical across all forward solvers, indicating that the discrepancy is associated with the backward KKT solve rather than the choice of forward optimizer.

For unregularized LPs, the $h$- and $b$-gradient panels exhibit nearly uniform agreement, with cosine similarities close to one for every solved configuration. The $q$-gradient panel is more solver dependent, particularly for well-conditioned LPs. For the sparse well-conditioned case, the available similarities range from $0.462$ for Gurobi to $0.929$ for OSQP. Moreover, full-KKT, reduced-KKT, and value function layer produce effectively identical $q$-gradient similarities for each solver-setting pair. This shows that the variation is not caused by the differentiation mode. The reason is explained below.

The LP behavior is consistent with non-unique primal solutions. For
\begin{equation*}
    \alpha(q)=\min_{x\in\mathcal F}q^T x,
\end{equation*}
the $q$-gradient corresponds to an optimal primal solution when the optimizer is unique. If several optimal solutions exist, different solvers may select different valid optimizers and hence different valid sensitivities (see Proposition~\ref{prop:subgradient_qbh}). The lower $q$-gradient cosine should therefore be interpreted primarily as optimizer-selection sensitivity rather than incorrect differentiation. The stable $h$- and $b$-gradient panels demonstrate that this ambiguity is concentrated in the primal sensitivity.

For LPs, the regularized variants add $\epsilon\|x\|_2^2/2$ with $\epsilon=10^{-3}$. Their $q$-gradient similarities lie between approximately $0.934$ and $0.995$, and all regularized LP configurations are solved successfully. This suggests that regularization improves the consistency of primal selection and numerical robustness.

\subsubsection{CPU Utilization and Computational Work}

\begin{figure*}[t]
    \centering
    \includegraphics[width=0.9\linewidth]{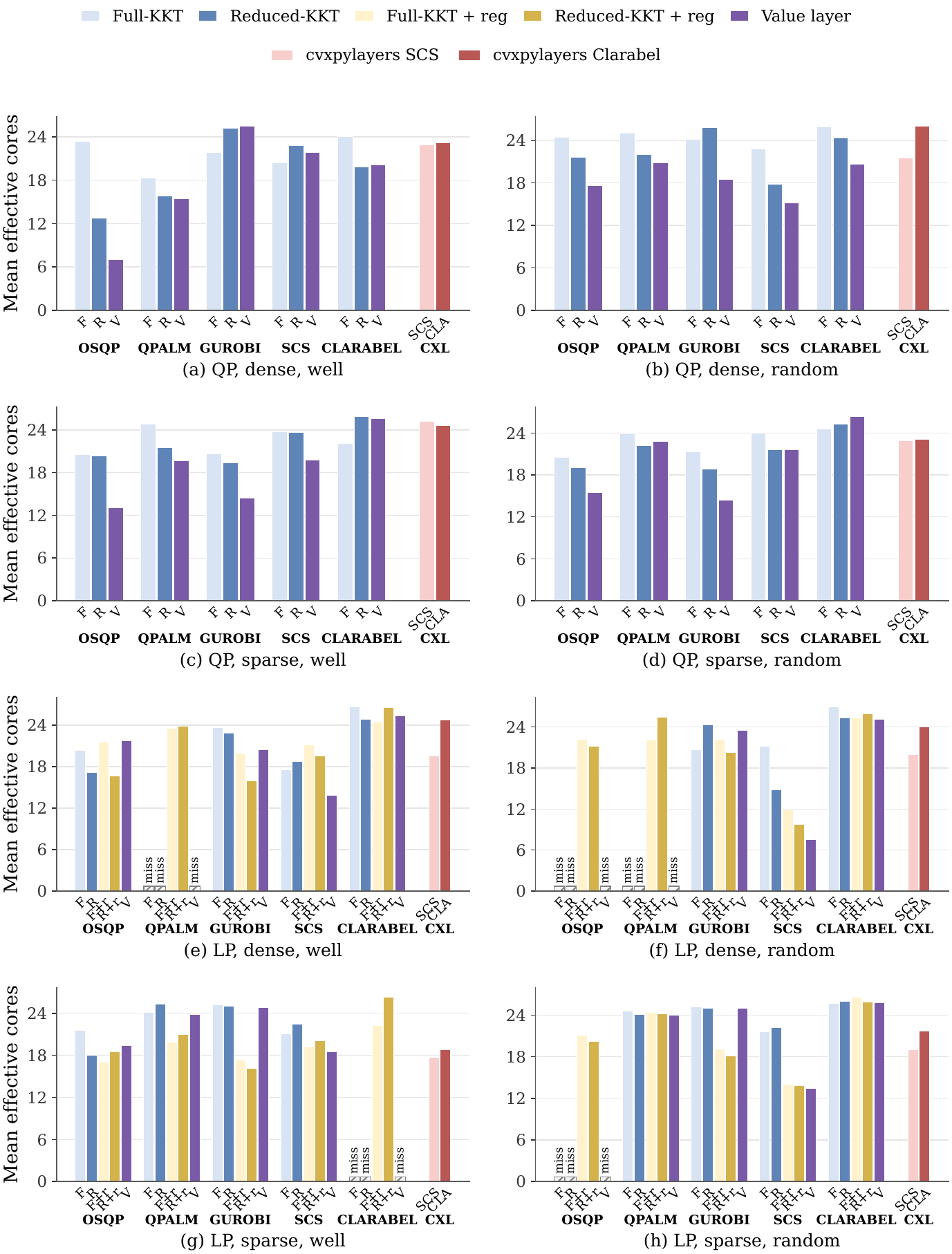}
    \caption{Mean effective CPU-core utilization, aggregated over the parent and worker processes during end-to-end forward and backward evaluation. DiffAPQP solution-map and value-function layers use combined warm start and problem-data update (WU); the CXL group reports the corresponding CvxpyLayers-SCS and CvxpyLayers-Clarabel references. }
    \label{fig:random_cpu}
\end{figure*}

\begin{figure*}[t]
    \centering
    \includegraphics[width=0.9\linewidth]{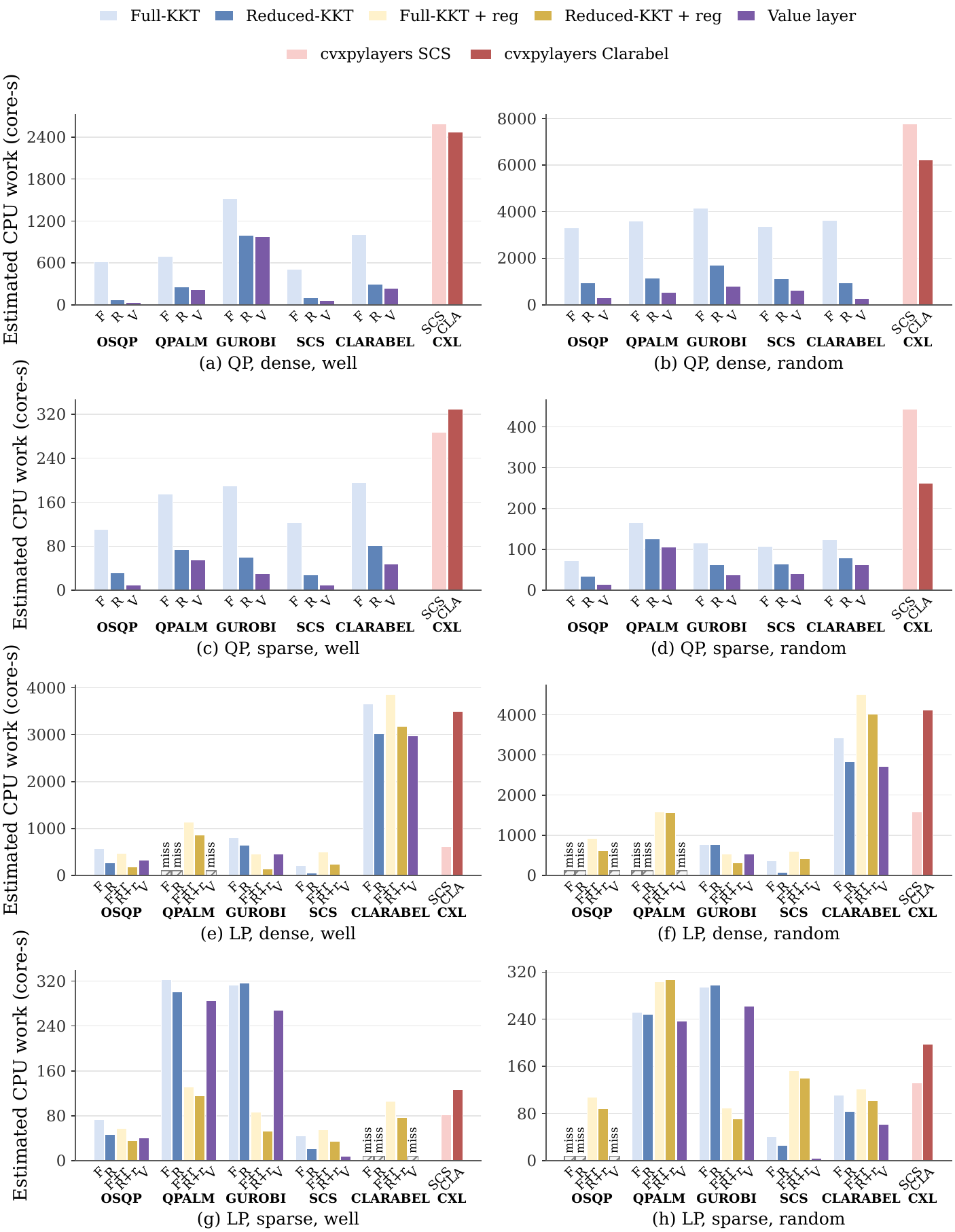}
    \caption{Estimated end-to-end CPU work in core-seconds, computed as the mean effective cores multiplied by the combined forward and backward runtime. DiffAPQP uses combined warm start and problem-data update (WU); the CXL group reports the corresponding CvxpyLayers-SCS and CvxpyLayers-Clarabel references.}
    \label{fig:random_cpu_work}
\end{figure*}

Fig.~\ref{fig:random_cpu} and Fig.~\ref{fig:random_cpu_work} report mean CPU utilization and estimated CPU work for the large benchmark problems. All experiments use batches of 32 and are restricted to 32 CPU cores according to Appendix~\ref{app:cpu_restriction}. CPU utilization includes the parent and recursive worker processes:
\begin{equation*}
    C_{{eff}}=\frac{\sum_p \Delta T_{{CPU},p}}
     {\Delta T_{{wall}}}
\end{equation*}
where $\Delta T_{{CPU},p}$ is the CPU time consumed by process $p$ during the measurement interval, including both user and system CPU time and $\Delta T_{{wall}}$ is the elapsed real-world time over the same measurement interval.
Thus, $C_{{eff}}=24$ means that approximately 24 cores were active on average.
\begin{tcolorbox}[
    colback=white,
    colframe=black,
    boxrule=0.5pt,
    left=4pt,
    right=4pt,
    top=4pt,
    bottom=4pt
]
The following observations are made:
\begin{enumerate}
    \item Both DiffAPQP and CvxpyLayers exploit substantial batch parallelism.
    \item Reduced-KKT uses slightly fewer effective cores than full-KKT but requires only about 36\% of its QP CPU work and 66\% of its LP CPU work.
    \item The value function layer is the most computationally economical option, requiring approximately 19\% and 43\% of full-KKT's CPU work for QPs and LPs, respectively.
    \item Sparse problems consume substantially fewer core-seconds than dense problems.
\end{enumerate}
\end{tcolorbox}

Most DiffAPQP and CvxpyLayers configurations use 18-26 effective cores, demonstrating substantial parallelism in both frameworks. For QPs, full-KKT, reduced-KKT, and the value function layer average 22.83, 21.31, and 18.80 cores, respectively, compared with 23.14 for CvxpyLayers-SCS and 24.24 for CvxpyLayers-Clarabel. For unregularized LPs, the corresponding DiffAPQP averages are 23.08, 22.41, and 20.82. Performance differences therefore cannot be attributed to either framework operating serially. Meanwhile, lower utilization does not necessarily imply lower efficiency. Full-KKT introduces substantial parallel backward computation, whereas reduced-KKT solves a smaller adjoint system and the value function layer largely avoids solution-map differentiation. We therefore estimate total processor consumption as
\begin{equation*}
    W_{{CPU}}=C_{{eff}}\cdot
\left(T_{{forward}}+T_{{backward}}\right)
\end{equation*}
measured in core-seconds.

For QPs, reduced-KKT requires approximately 36\% of full-KKT's CPU work in geometric-mean terms, while the value function layer requires only 19\%. For LPs, these ratios are approximately 66\% and 43\%, respectively. Regularized reduced-KKT consumes approximately 69\% of the CPU work required by regularized full-KKT. These reductions show that the lower utilization of reduced-KKT and the value function layer reflects less computation rather than limited parallelism.

Sparsity also substantially reduces CPU work. Dense and sparse problems often use similar numbers of effective cores, but sparse configurations finish sooner and consume considerably fewer core-seconds. Solver choice has a similar effect. Clarabel frequently reaches 25-27 effective cores, particularly for dense LPs, but can still consume substantial CPU work because of its longer forward pass. 

Overall, both frameworks exploit multicore execution effectively. Reduced-KKT provides a strong balance between parallelism and processor consumption, while the value function layer is the most economical option when only optimal-value sensitivities are required. Mean effective cores should be interpreted as parallel utilization, whereas core-seconds more directly represent computational efficiency.

\subsubsection{Memory Usage}

\begin{figure*}[t]
    \centering
    \includegraphics[width=0.9\linewidth]{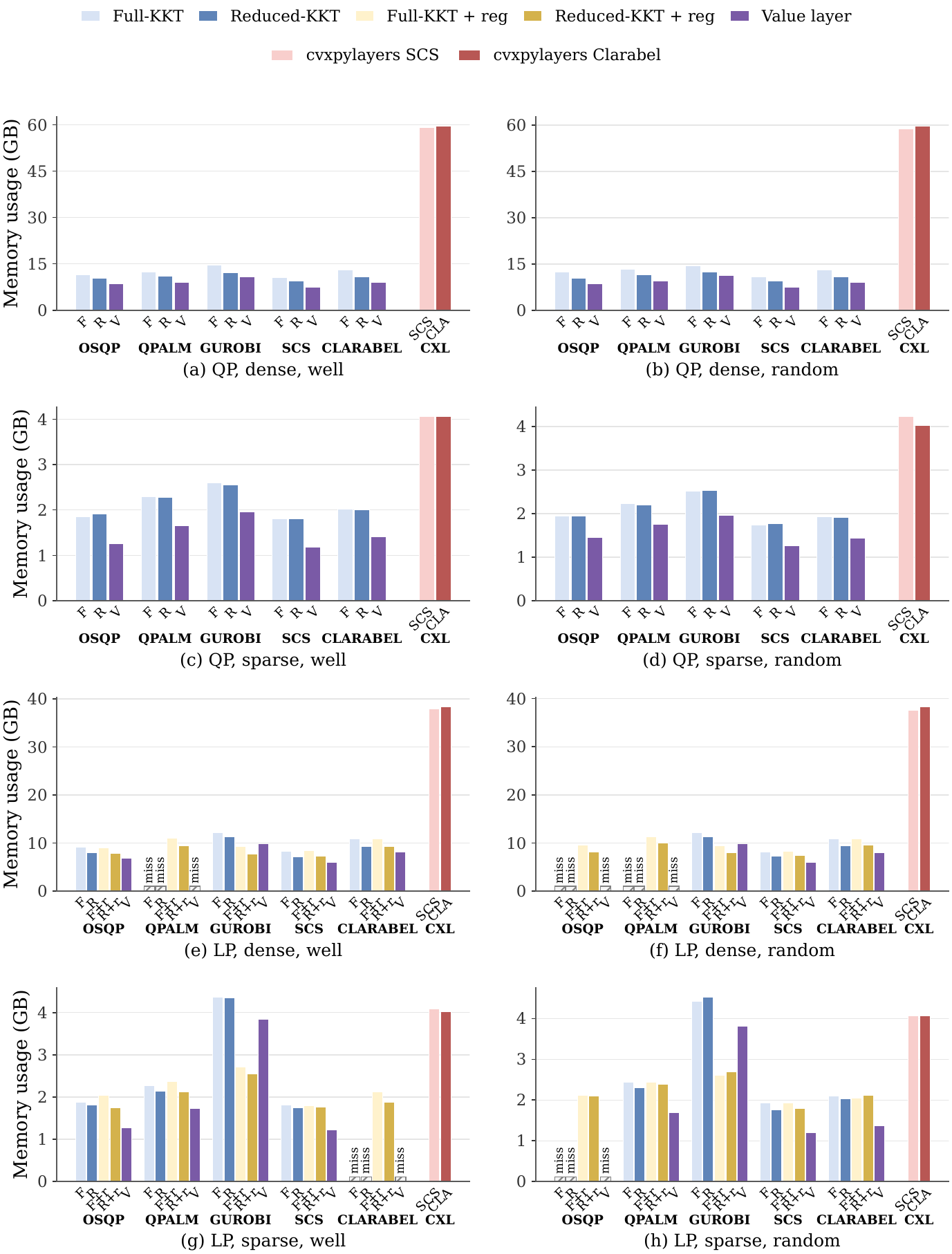}
    \caption{Peak increase in system-used memory relative to the pre-construction benchmark baseline, measured during layer initialization and end-to-end evaluation. DiffAPQP uses combined warm start and problem-data update (WU); the CXL group reports the corresponding CvxpyLayers-SCS and CvxpyLayers-Clarabel measurements.}
    \label{fig:random_memory}
\end{figure*}

Fig.~\ref{fig:random_memory} reports the peak increase in system-used memory from starting time $t_0$,
\begin{equation*}
    \Delta M_{{peak}} = \max_t M_{{sys}}(t)-M_{{sys}}(t_0)
\end{equation*}
According to Appendix~\ref{app:usage_reporting}, memory sampling begins before layer construction and therefore includes worker creation, solver workspaces, the DiffAPQP base solve and cache, and forward and backward execution. Unsolved configurations are excluded.
\begin{tcolorbox}[
    colback=white,
    colframe=black,
    boxrule=0.5pt,
    left=4pt,
    right=4pt,
    top=4pt,
    bottom=4pt
]
The following observations are made:
\begin{enumerate}
    \item With the same SCS or Clarabel backend, DiffAPQP reduces peak memory increase by 72\%-87\% for dense problems relative to CvxpyLayers;
    \item For sparse problems, the corresponding DiffAPQP memory reduction is approximately 48\%-71\%;
    \item Reduced-KKT uses about 7\%-9\% less memory than full-KKT;
    \item The value function layer provides the largest internal saving.
\end{enumerate}
\end{tcolorbox}

DiffAPQP requires substantially less memory than CvxpyLayers under the same solver backend. For dense QPs with SCS, CvxpyLayers requires approximately 59 GB, compared with 10.6-10.9 GB for full-KKT, 9.4 GB for reduced-KKT, and 7.4-7.5 GB for the value function layer. These represent reductions of approximately 82\%, 84\%, and 87\%. With Clarabel, CvxpyLayers requires approximately 60 GB, whereas the corresponding DiffAPQP variants require approximately 13.0, 10.8, and 9.0 GB, giving reductions of 78\%-85\%.

The same pattern holds for dense LPs. CvxpyLayers requires approximately 38 GB with either backend. DiffAPQP-SCS uses approximately 8.2, 7.2, and 5.9 GB for full-KKT, reduced-KKT, and value function layer, respectively. DiffAPQP-Clarabel uses approximately 10.9, 9.3, and 8.1 GB. Overall, DiffAPQP reduces dense-LP memory by approximately 72\%-84\% relative to CvxpyLayers with the same solver.

Sparse matrices reduce memory requirements for both frameworks. CvxpyLayers uses approximately 4.0-4.2 GB for sparse problems, while DiffAPQP generally uses 1.7-2.1 GB with the solution-map variants and 1.2-1.4 GB with the value function layer. This corresponds to reductions of approximately 48\%-57\% for full- and reduced-KKT and 65\%-71\% for value function layer. The larger advantage on dense problems suggests that CvxpyLayers incurs substantial overhead from dense canonical or derivative data during batched execution.

Within DiffAPQP, reduced-KKT uses approximately 7\%-8\% less memory than full-KKT for unregularized problems and approximately 9\% less for regularized LPs. The value function layer provides the largest saving, using approximately 28\% less memory for QPs and 26\% less for LPs than full-KKT.

To conclude, the consistent same-backend differences provide strong evidence that DiffAPQP is substantially more memory efficient than CvxpyLayers.



\FloatBarrier
\clearpage
\IEEEpeerreviewmaketitle
\bibliographystyle{IEEEtran}
\bibliography{IEEEabrv,Reference.bib}

\bstctlcite{IEEEexample:BSTcontrol}


\ifCLASSOPTIONcaptionsoff
  \newpage
\fi
\end{document}